\documentstyle[11pt,fps]{article}
\makeatletter
\@addtoreset{equation}{section}
\makeatother

\begin{document}

\begin{center}

{\LARGE\bf The Scaling of Exact and Approximate \\  \ \\
Ginsparg-Wilson Fermions}

\vspace*{1.5cm}

{\large W. Bietenholz} \\ \ \\ 
NORDITA \\
Blegdamsvej 17 \\
DK-2100 Copenhagen \O , Denmark \\ 
\ \\
{\em and} \\
\ \\
{\large I. Hip} 
\footnote{Supported by Fonds zur F\"{o}rderung der Wissenschaftlichen
Forschung in \"{O}sterreich, \\ \hspace*{5mm} Project P11502-PHY.} \\ \ \\
Institut f\"{u}r Theoretische Physik \\
Universit\"{a}t Graz \\
A-8010 Graz, Austria \\

\ \\
\ \\

Preprint \ \ NORDITA-99/10-HE \\ 
\hspace*{8mm} hep-lat/9902019
%, \ UNIGRAZ-UTP-02-14-99

\vspace*{1cm}

\end{center}

We construct a number of lattice fermions,
which fulfill the Ginsparg-Wilson relation either exactly
or approximately, and test them in the framework of the
2-flavor Schwinger model. We start from explicit
approximations within a short range, and study this
formulation, as well as its correction to an exact
Ginsparg-Wilson fermion by the ``overlap formula''.
Then we suggest a new method to realize this correction
perturbatively, without using the tedious square root
operator. In this way we combine many favorable properties:
good chiral behavior, small mass renormalization,
excellent scaling and rotational invariance,
as well as a relatively modest computational effort, 
which makes such formulations most attractive for QCD.

\newpage

\section{Introduction}

Recently there have been intensive new activities
to circumvent the notorious Nielsen-Ninomiya No-Go theorem \cite{NN}
for chiral fermions on the lattice. That theorem has a rather complicated
set of minimal assumptions; we simplify them slightly to the following
statement: for an (undoubled) lattice fermion, unitarity, 
discrete translation invariance, locality and (full) chiral symmetry 
cannot coexist. To get around this theorem, the breaking of each of these 
properties has been tried. Examples are (referring to the above order)
one-sided lattice Dirac operators, random
lattices, SLAC and Rebbi fermions, and finally the Wilson fermion.

The latter breaks the full chiral symmetry in a rather hard way,
destroying essential physical properties related to chirality.
(By full chiral symmetry we mean the relation 
$\{ D , \gamma_{5} \} = 0$, where $D$ is the lattice Dirac operator,
and the curly bracket denotes the anti-commutator.)
The subject of this paper is a recently re-discovered approach
to perform such a chiral symmetry breaking in a much softer
way, preserving a modified but continuous form of chiral
symmetry at finite lattice spacing.

That remnant lattice chiral symmetry transformation has first been 
written down in Ref.\ \cite{ML}. Its generalized form reads
(we use a short-hand notation for the convolutions in coordinate space)
\begin{equation}
\bar \psi_{x}  \to  \Big( \bar \psi \ 
(1 + \epsilon \, [1-DR \, ] \gamma_{5}) \Big)_{x} \ , \quad
\psi_{x}  \to  \Big( (1 + \epsilon \, [1-DR \, ] \gamma_{5}) \ 
\psi \Big)_{x}
\end{equation}
where $R$ is a {\em local} Dirac scalar, i.e.\ it decays at 
least exponentially. We now require invariance of the fermionic
Lagrangian \ $\bar \psi D \psi$ \ to $O(\epsilon )$. This amounts
to the condition
\begin{equation}
\{ D_{x,y} , \gamma_{5} \} = 2 (D \gamma_{5} R D)_{x,y} \ ,
\end{equation}
which is known as the Ginsparg-Wilson relation (GWR).
The crucial point is that the term $R$, which describes
the chiral symmetry breaking of $D^{-1}$,
\begin{equation}
R_{x,y} = \frac{1}{2} \gamma_{5} \{ D^{-1}_{x,y} , \gamma_{5} \} \ ,
\end{equation}
is local, which is not the case for Wilson fermions, massive fermions
etc. Therefore the pole structure of $D^{-1}$ is not affected by $R$.

Ginsparg and Wilson pointed out a long time ago
that this is a particularly soft way to break chiral symmetry
on the lattice \cite{GW}. 
Renewed interest was attracted to this approach especially by 
Ref.\ \cite{HLN}. It was demonstrated that it preserves
the triangle anomaly \cite{GW},
that it avoids additive mass renormalization and mixing of
matrix elements \cite{Has}, and that it reproduces the soft
pion theorems \cite{SC}.
So far we refer to vector theories, but the GWR
even serves as a basis for the construction of chiral
gauge theories \cite{MLcg}. As further applications, the GWR
provides for instance a safe continuum limit of the chiral
anomaly \cite{anomal} and of the spontaneous chiral
symmetry breaking \cite{GB}.
It also opens new perspectives in other fields, like
random matrix models \cite{random}.

As a general ansatz for a solution of the GWR, we write
\begin{equation}
D^{-1} = D^{-1}_{\chi} + R \ , \quad {\rm where} \ \
\{ D^{-1}_{\chi},\gamma_{5} \} = 0 \ .
\end{equation}
Now a suitable operator $D_{\chi}$ --- with the correct continuum
limit --- has to be identified, which is non-trivial even in the 
free case. If it suffers from doubling, non-unitarity
or SLAC type non-locality (finite gaps
of the free $D_{\chi}(p)$, where $p$ is the momentum), 
then that disease is inherited by $D$.
Hence all these options must be discarded.
A way out is, however, a non-locality of the Rebbi type \cite{Rebbi},
where $D_{\chi}(p)$ has divergences.
This still allows for locality of $D$.

That is exactly the mechanism which is at work to yield
local perfect fermions \cite{GW} (lattice fermions
without lattice artifacts). Here the term $R^{-1}$
plays a specific r\^{o}le: in the block variable renormalization
group transformation, which leads asymptotically to a perfect action,
it is the kernel between the blocks in a Gaussian transformation
term. \footnote{Only the limit $R \to 0$ 
(``$\delta$ function blocking'') leads to non-locality.}
Also in the interacting case the perfect fermion is a solution
of the GWR, in agreement with the fact that it breaks the chiral
symmetry only superficially (in the manifest form of the lattice
action), but not with respect to the physical observables \cite{FPA}.
However, the perfect fermion can only be constructed perturbatively
or in the classical approximation (fixed point action, FPA) 
\cite{FPA,QuaGlu,LP}.
It turned out that the latter is a solution of the GWR too \cite{HLN},
and its locality is optimized by the choice
\begin{equation} \label{standGWR}
R_{x,y} = \frac{1}{2} \delta_{x,y} \ , \quad
\{ D_{x,y},\gamma_{5} \} = (D \gamma_{5} D)_{x,y} \ .
\end{equation}
The use of a FPA, together with the corresponding classically perfect
topological charge, guarantees that the index theorem is correctly 
represented on the lattice \cite{HLN}.

We refer to eq.\ (\ref{standGWR}) as the ``{\em standard form}'' 
of the GWR.
For a resulting FPA, the index theorem has been confirmed
numerically in the Schwinger model \cite{FL}, where the topological
charge was also defined in the spirit of the FPA.

Except for the FPA, another GW fermion was discovered by
H.\ Neuberger \cite{Neu}. The locality of that solution has been
established analytically in a smooth gauge background
and numerically up to moderate coupling strength in QCD \cite{HJL}. 

The relaxation of the full chiral symmetry
condition to the GWR allows the fermions to be local,
but it has been conjectured that they can still not
be ``ultralocal'' \cite{WB}. This means that their couplings may decay 
exponentially, but they cannot stop at a finite number of lattice 
spacings, not even in the free case. 
In fact, this has been demonstrated for the
standard form of the GWR \cite{IH}
and for a more general class of GW kernels $R$ 
in any dimension $d \geq 2$ \cite{WBultra}.

Against this background, we construct in the next Section
lattice fermions, which satisfy the GWR at least 
to a good approximation, but which have couplings only in a short 
range, so that they can be implemented directly for simulations.
\footnote{There is some conceptual --- though not technical ---
similarity with the papers in Ref.\ \cite{TdGchi}.}
At the same time, we optimized other essential properties, in particular
the scaling behavior and the approximate rotation invariance.
We emphasize that the GWR does not guarantee
a good quality of those properties. In fact, it has been
observed \cite{FHL,FHLW} --- and it will be confirmed in Section 3 --- 
that the Neuberger fermion does a rather poor job with that
respect.

In addition we do our best to keep the computational effort
modest. In the framework of the Schwinger model,
where our study takes place, we can also simulate
very complicated actions. 
%(a striking example is the FPA from Ref.\ \cite{LP} mentioned before).
However, we are interested in working out a formulation,
which has the potential to be carried on and applied to $d=4$.
All the above properties (in particular excellent chirality and 
scaling) are also manifest in the FPA for the Schwinger model
in Ref.\ \cite{LP}, except for the simplicity. 
That (approximate) FPA involves 123 independent couplings --- 
and all together 429 terms per site.
Unfortunately this makes an analogous formulation in $d=4$
inapplicable. So far, all attempts to construct a
useful approximate FPA for QCD got stuck in the fermionic part 
of the action, hence it is strongly motivated to search for
a simplified alternative with similar qualities.
%, in particular concerning chirality {\em and} scaling.

\section{Hypercubic approximate Ginsparg-Wilson \\ fermions}

To improve the lattice fermion, i.e.\ to suppress lattice
artifacts, we have to include couplings 
to lattice sites beyond nearest neighbors. However, the
systematic extension to various lattice spacings amplifies
the number of couplings very rapidly.
%so that one has to economic with the non-standard terms.
As an option, which seems to allow for a powerful improvement,
but which is still tractable in QCD simulations \cite{TdGHF,SESAM},
we focus on the ``{\em hypercube fermion}'' (HF), where
$\bar \psi_{x}$ is coupled to $\psi_{y}$ if
$\vert x_{\mu} - y_{\mu} \vert \leq 1$ for $\mu = 1\dots d$,
i.e.\ we couple all sites inside a $d$ dimensional unit
hypercube.

\subsection{Free hypercube fermions}

For free fermions, perfect actions can be computed 
analytically \cite{QuaGlu}. However, they can only be local
in the sense that their couplings decay exponentially,
in agreement with the conjecture about the absence of 
ultralocal GW fermions. For practical
purposes the parameters in the renormalization
group transformation can be tuned so that the decay 
becomes very fast. This is the case for the parameters
corresponding to the standard GWR.
There the couplings were truncated
to the unit hypercube by means of periodic boundary conditions
over three lattice spacings \cite{MIT}.
The resulting {\em truncated perfect} HF (TP-HF) has strongly
improved scaling properties compared to the Wilson fermion.
Since the truncation is only a small modification, it also
approximates the standard GWR to a good accuracy \cite{WB}.

To fix our notation, we write the free lattice Dirac operator as
\begin{equation} \label{freefermi}
D_{x,x+r} = D (r) = \rho_{\mu}(r) \gamma_{\mu} + \lambda (r) \ ,
\quad (x,r \in Z \!\!\! Z ^{d} ) \ ,
\end{equation}
where we assume the sensible symmetries:
$\rho_{\mu}$ is odd in the $\mu$ direction and even
in all other directions, while the Dirac scalar $\lambda$
is entirely even. In addition $\rho_{\mu}$ is invariant
under permutations of the non-$\mu$ axes, and $\lambda$
under any permutation of the axes.
Of course $D$ must also have the correct continuum limit,
$D(p) = i p_{\mu} \gamma_{\mu} + O(p^{2})$.

As a measure for the total violation of the free GWR, we sum
the squared violations in each site,
\begin{eqnarray}
{\cal V} &=& \sum_{r} \Big[ \sum_{x,y,z} 2 D(x) \gamma_{5} R(y) D(z) 
\delta_{r,x+y+z} - \{ D(r), \gamma_{5} \}  \Big] ^{2} \nonumber \\
&=& 4 \sum_{r} \Big[ \sum_{x,y,z} [ \lambda (x) R(y) \lambda (z)
- \rho_{\mu}(x) R(y) \rho_{\mu}(z) ] \delta_{r,x+y+z} 
- \lambda (r) \Big] ^{2} \ .
\end{eqnarray}
For the standard GWR (\ref{standGWR}) this simplifies to
\begin{equation}
{\cal V}_{st} =  \sum_{r} \Big[ \sum_{x} [ \lambda (x) \lambda (r-x)
- \rho_{\mu}(x) \rho_{\mu}(r-x) ] - \lambda (r) \Big]^{2} \ .
\end{equation}
If we just optimize the couplings so that ${\cal V}_{st}$
becomes minimal, then we can still do somewhat better
than the TP-HF. For $d=2$ this can be seen from Table \ref{tabHF},
where we denote that ``{\em chirally optimized}\,'' HF as CO-HF.
However, the chiral optimization makes the scaling
behavior a little worse than it is the case for the
TP-HF, see below. On the other hand,
if we search for excellent scaling, then we end up
with a HF that we call SO-HF ({\em scaling optimized}\,), and which is also
included in Table \ref{tabHF}. Its value for ${\cal V}_{st}$ is still 
small, but clearly larger than in the previous two cases. The
TP-HF can therefore be seen as a compromise between the two
optimizations with respect to just one property.
Note that the couplings are similar in all these three HFs.

\begin{table}
\begin{center}
\begin{tabular}{|c|c|c|c|}
\hline
 & TP-HF & CO-HF & SO-HF \\
\hline
\hline
$\rho^{(1)} := \rho_{1}(1,0)$ &  ~0.30938846 &  ~0.30583220 & 0.334 \\
\hline
$ \rho^{(2)} := \rho_{1}(1,1)$ &  ~0.09530577 &  ~0.09708390 & 0.083 \\
\hline
$\lambda_{0} := \lambda (0,0)$ &  ~1.48954496 &  ~1.49090692 & 1.5   \\
\hline
$\lambda_{1} := \lambda (1,0)$ & --0.24477248 & --0.24771369 & --0.25 \\
\hline
$ \lambda_{2} := \lambda (1,1) $ & --0.12761376 & --0.12501304 & --0.125 \\
\hline
\hline
${\cal V}_{st} $ & $3.008 \cdot 10^{-4}$ &  $1.006 \cdot 10^{-4}$ & 
$65.518 \cdot 10^{-4}$\\
\hline
\hline
\end{tabular}
\end{center}
\caption{\sl The couplings of free hypercube fermions ---
truncated perfect, chirally optimized and scaling optimized ---
and their violations of the standard GWR.
Note the constraints $\rho^{(1)}+ 2 \rho^{(2)}=1/2$ and
$\lambda_{0} + 4 ( \lambda_{1} + \lambda_{2}) =0$ from the
continuum limit and from mass zero, respectively.}
\label{tabHF}
\end{table}

Actually the comparison of ${\cal V}_{st}$ is not really fair for the 
SO-HF, because that fermion is not necessarily related to the standard GWR.
If we allow for $R_{x,y} = r_{0} \delta_{x,y}$  ($r_{0}$ arbitrary),
then its value ${\cal V}$ drops to $18.413 \cdot 10^{-4}$ 
(for $r_{0}=0.48950632$). If $R$ is generalized further to an even
hypercubic form (like $\lambda$), we arrive at
${\cal V}= 15.977 \cdot 10^{-4}$, but this
is still not in the same order of magnitude as the TP-HF and CO-HF.

We can also perform the minimization of the HF couplings
for such generalized GW kernels $R$. 
However, we have to avoid the trivial
minimum at $\lambda = R=0$ (naive fermion), so we insert
another constraint on $\lambda$. We require the mapping 
to $d=1$ to reproduce the 1d Wilson fermion, 
$\lambda_{1}+ 2 \lambda_{2} =-1/2$ (which is also the
case for TP-HF). The minimum is still found in the same vicinity,
$\rho^{(2)} = 0.09909187$, $\lambda_{2} = -0.13035191$, and
$R(0,0)=0.50514453$, $R(1,0)=-0.00141375$, $R(1,1)=0.00183004$,
which shows that --- at least with this extra constraint ---
the vicinity of the standard GWR plays indeed a special
r\^{o}le. The violation then amounts to 
${\cal V} = 0.793 \cdot 10^{-4}$,
so we can apparently not proceed to lower orders of magnitude
any more. \\

\begin{figure}[hbt]
   \begin{tabular}{ccc}
      \hspace{-0.5cm}
\def\fpsangle{270} \epsfxsize=39mm \fpsbox{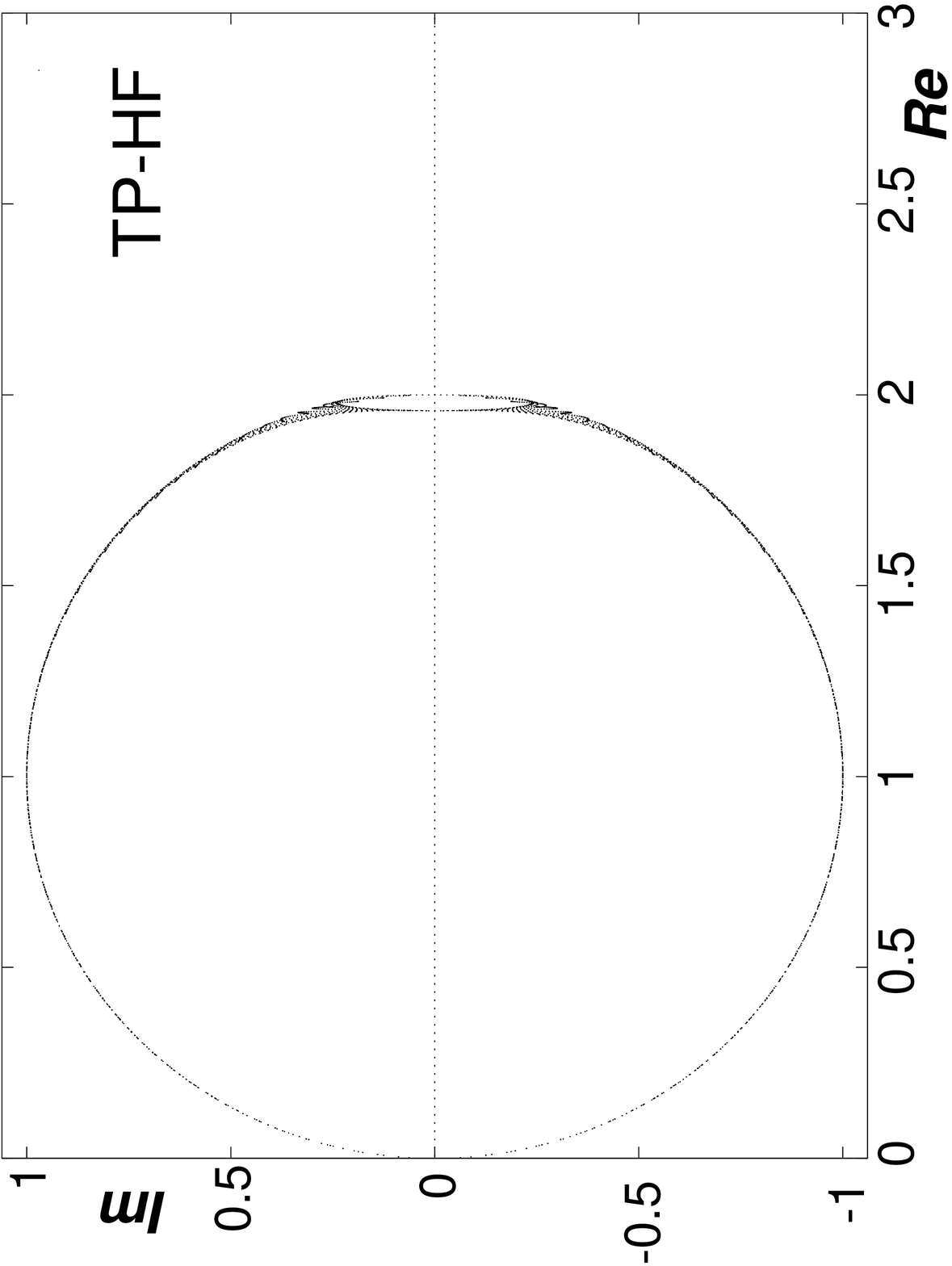} &
      \hspace{-0.55cm}
\def\fpsangle{270} \epsfxsize=39mm \fpsbox{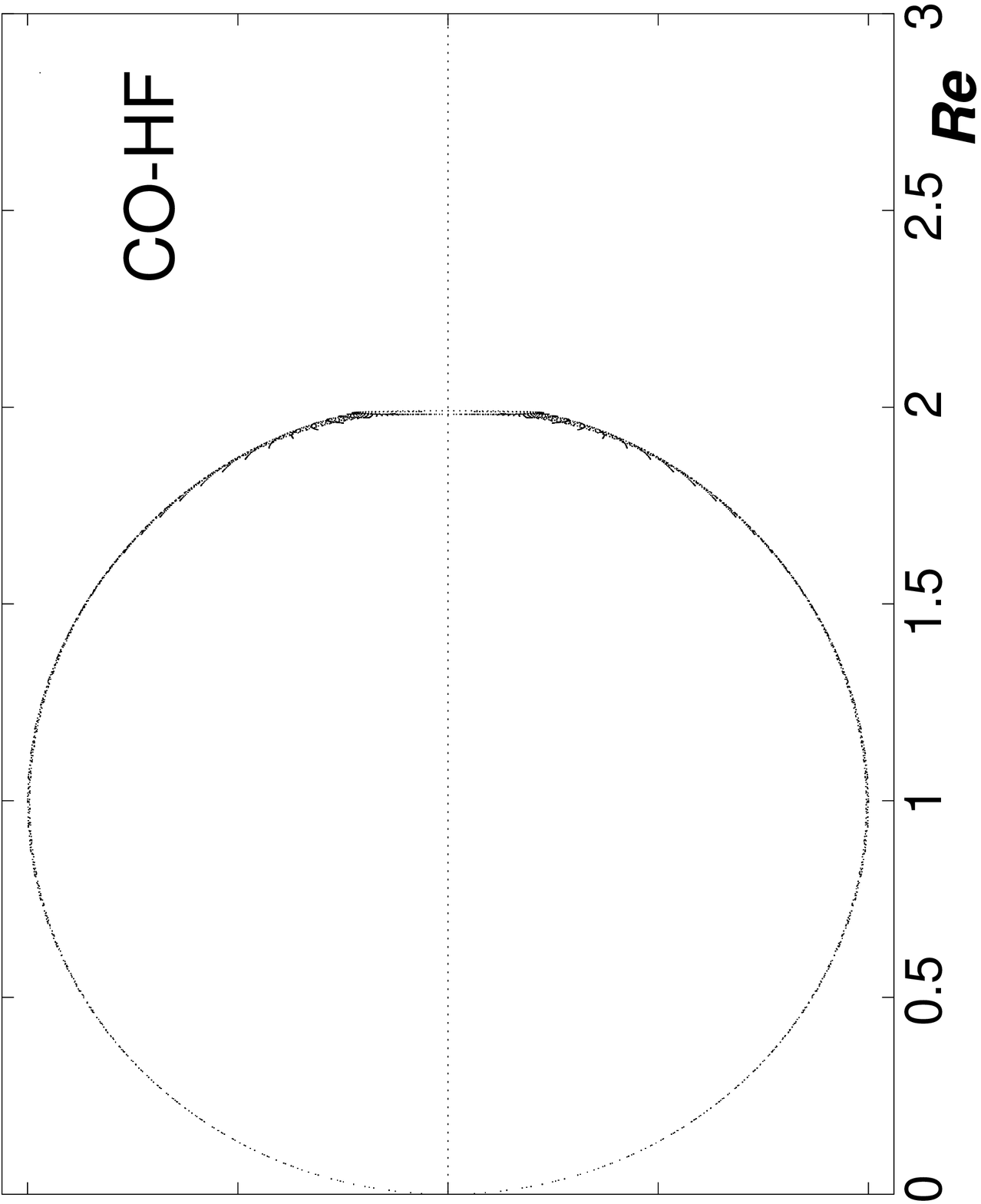} &
      \hspace{-0.55cm}
\def\fpsangle{270} \epsfxsize=39mm \fpsbox{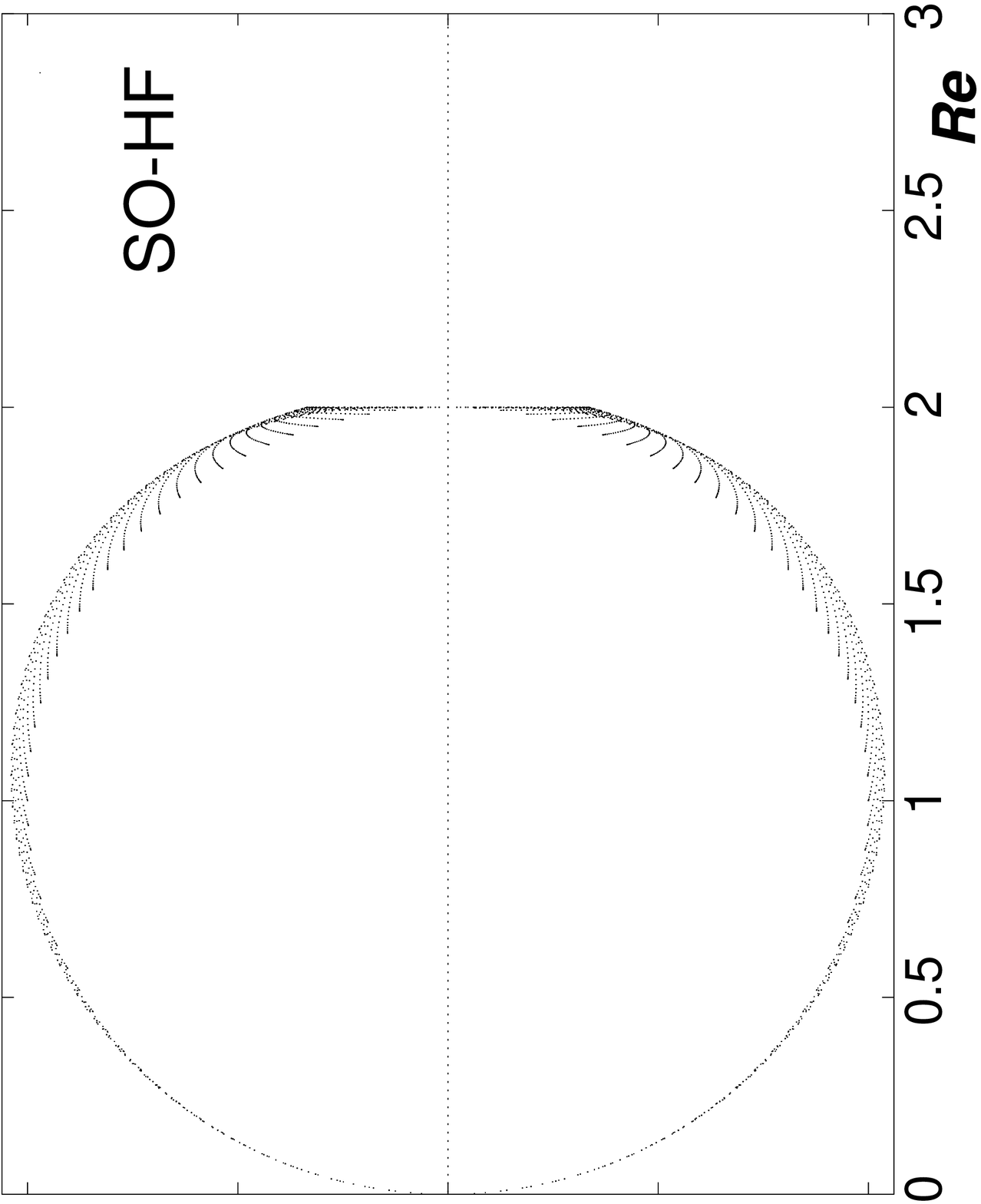}
   \end{tabular}
   \caption{\sl The free spectra of 3 hypercube fermions:
truncated perfect (left), chirally optimized (center) and
scaling optimized (right), on a $100\times 100$ lattice.}
\label{figspectrumHF}
\end{figure}

Hence we return to the form $R_{x,y}=r_{0} \delta_{x,y}$. In this case,
there is a simple way to illustrate the accuracy of the GWR. 
For an exact GW fermion, the spectrum lies on the circle 
in the complex plane with center and radius $1/(2r_{0})$ \cite{HLN}. 
This holds with or without gauge interaction.
Therefore we can just check how close
the eigenvalues of the approximate GW fermion are to that
circle. This is shown in Fig.\ \ref{figspectrumHF}, 
which confirms the hierarchy measured from ${\cal V}$.
The strongest deviations occur in the arc opposite to zero.
This arc corresponds to high momenta and therefore a fine
resolution, which is sensitive to small inaccuracies of the GWR.
%[measure spectral deviation also quantitatively]

\begin{figure}[hbt]
%\begin{center}
\hspace*{25mm}
\def\fpsangle{270}
\epsfxsize=70mm
\fpsbox{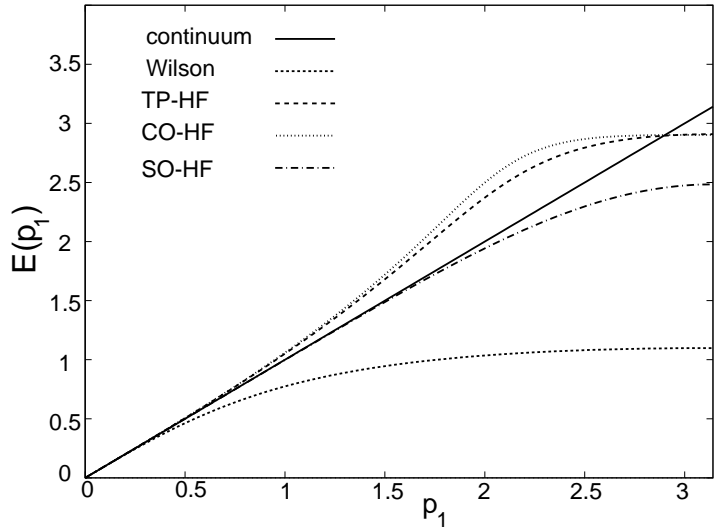}
%\end{center}
\vspace{-3mm}
\caption{\sl The dispersion relation for free hypercube fermions
compared to the Wilson fermion and to the continuum.
(The energies of all higher branches keep above 4.4.)}
%\vspace{1mm}
\label{figfreedisp}
\end{figure}

However, as we emphasized in the introduction, we do not want
to concentrate solely on the chiral quality. In particular
we want to consider the scaling behavior as well, and
we now take a first look at the impact of such a GWR optimization
with that respect. Fig.\ \ref{figfreedisp} shows the dispersion relations 
for the free HFs mentioned before, and we see that all the three
are strongly improved compared to the Wilson fermion. We also
see that the hierarchy among the HFs is
inverted. This observation motivated the consideration
of the SO-HF, which was constructed by hand. (Some comments on the 
construction are given in the appendix.)

\begin{figure}[hbt]
\begin{tabular}{cc}
\hspace{-0.6cm}
\def\fpsangle{270} \epsfxsize=55mm \fpsbox{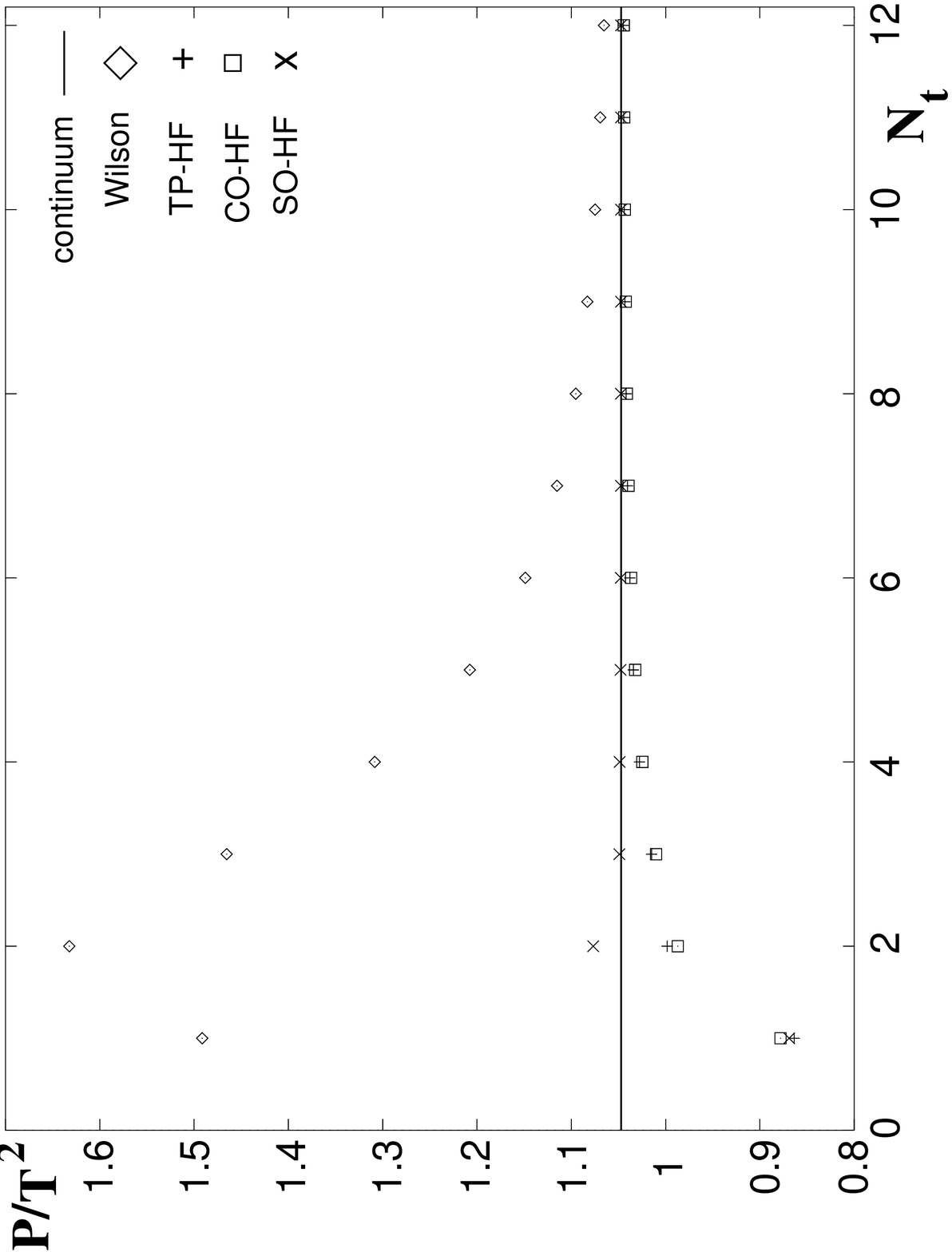} &
\hspace{-5mm}
\def\fpsangle{270} \epsfxsize=55mm \fpsbox{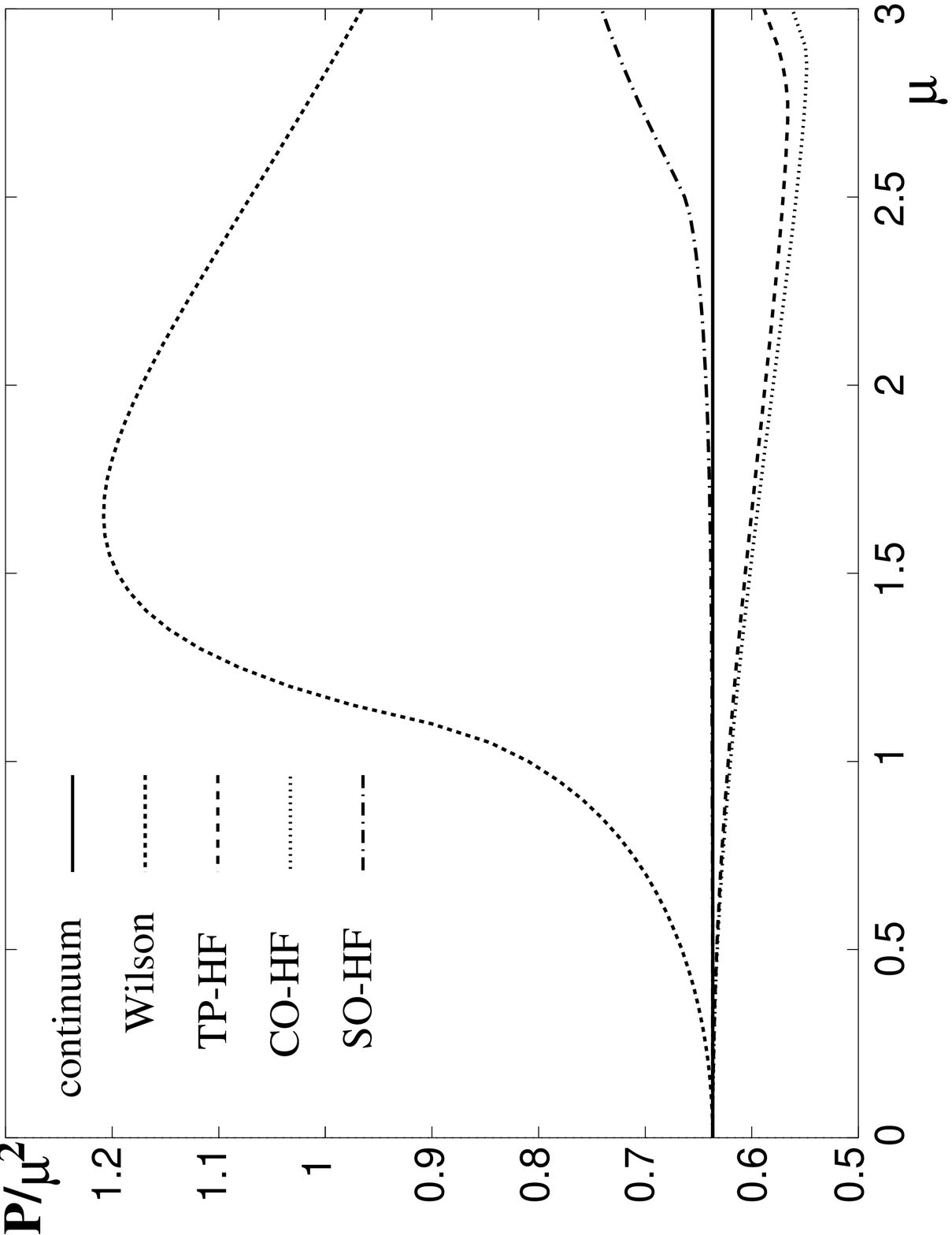}
\end{tabular}
\caption{\sl Thermodynamic scaling ratios.
Left: The ratio $P/T^{2}$ (at $\mu_{c}=0$) as a function of the number
$N_{t}$ of lattice points in the Euclidean time direction.
Right: The ratio $P/\mu_{c}^{2}$ at $T=0$ as a function of the
chemical potential $\mu_{c}$.}
\label{figthermo}
\end{figure}

We also want to compare thermodynamic scaling properties of the
free HFs. In Fig.\ \ref{figthermo} we show the
ratios $P/T^{2}$ at $\mu_{c} =0$, resp. $P/\mu_{c}^{2}$ at $T=0$,
which are scaling quantities in $d=2$ ($P$: pressure, $T$: temperature,
$\mu_{c}$: chemical potential). In the HF actions, $\mu_{c}$ is 
incorporated according to the prescription in Ref.\ \cite{chem}.
For $T \to 0$ (many lattice points $N_{t}$
in Euclidean time direction), resp. $\mu_{c} \to 0$, all HFs converge to
the continuum ratio, but the speed of convergence differs strongly.
The relative quality of the scaling behavior, which was suggested
by the dispersion, is confirmed. Again the SO-HF looks particularly 
impressive.

%\begin{figure}[hbt]
%\begin{center}
%%\hspace{20mm}
%\def\fpsangle{270}
%\epsfxsize=70mm
%\fpsbox{pres.eps}
%\end{center}
%\vspace{-6mm}
%\caption{\sl The ratio $P/T^{2}$ (at $\mu_{c}=0$) as a function of the number
%$N_{t}$ of lattice points in the Euclidean time direction.}
%\label{figpres}
%\end{figure}

%\begin{figure}[hbt]
%\begin{center}
%%\hspace{20mm}
%\def\fpsangle{270}
%\epsfxsize=70mm
%\fpsbox{presmu.eps}
%\end{center}
%\vspace{-6mm}
%\caption{\sl The ratio $P/\mu_{c}^{2}$ at $T=0$ as a function of the
%chemical potential $\mu_{c}$.}
%%\vspace{1mm}
%\label{figpresmu}
%\end{figure}

%% construction: TP-HF, CO-HF, SO-HF (scaling [or spectrum] optimal, 
%% justification below) \\
%% Table for couplings and total GWR violation\\
%% mention further GW optimizations \\
%% special r\^{o}le of standard GWR, spectrum as indicator. \\
%% Impact on other properties:
%% free dispersion relations, thermodynamics: $P/T^{2}$, $P/\mu^{2}$.

\subsection{Applications to the Schwinger model}

We now proceed to the 2-flavor Schwinger model (QED$_{2}$), and we
attach the free fermion couplings equally to the shortest
lattice paths only. Moreover we add a clover term, which turned
out to be useful, and we fix its coefficient to $c_{SW}=1$.
 
The classically perfect fermion-gauge vertex function
includes a number of plaquette couplings spread
over some range \cite{MIT}.
However, if we sum them up and ``compactify'' them all into
the clover term, we obtain $c_{SW}=1$. This holds 
for the ``compactified'' on-shell $O(a)$ improvement of 
any of our massless HFs.
Note that --- unlike QCD --- this value does not 
get renormalized due to the super-renormalizability of the 
Schwinger model \cite{clover2}.
Therefore, $c_{SW}=1$ provides a non-perturbative $O(a)$
improvement for the Wilson fermion as well as the HFs.

We use quenched configurations on a $16 \times 16$ lattice.
However, in the evaluation of dispersion relations and
correlation functions (see below) the square of the
determinant is included as a weight factor, following
the prescription in Ref.\ \cite{FHL}, Section 3.
For the pure gauge part, we use here and throughout this
paper the Wilson plaquette action, which is actually perfect
in 2d Abelian gauge theory \cite{QuaGlu}.

\begin{figure}[hbt]
\begin{tabular}{cc}
\hspace{-0.6cm}
\def\fpsangle{270} \epsfxsize=55mm \fpsbox{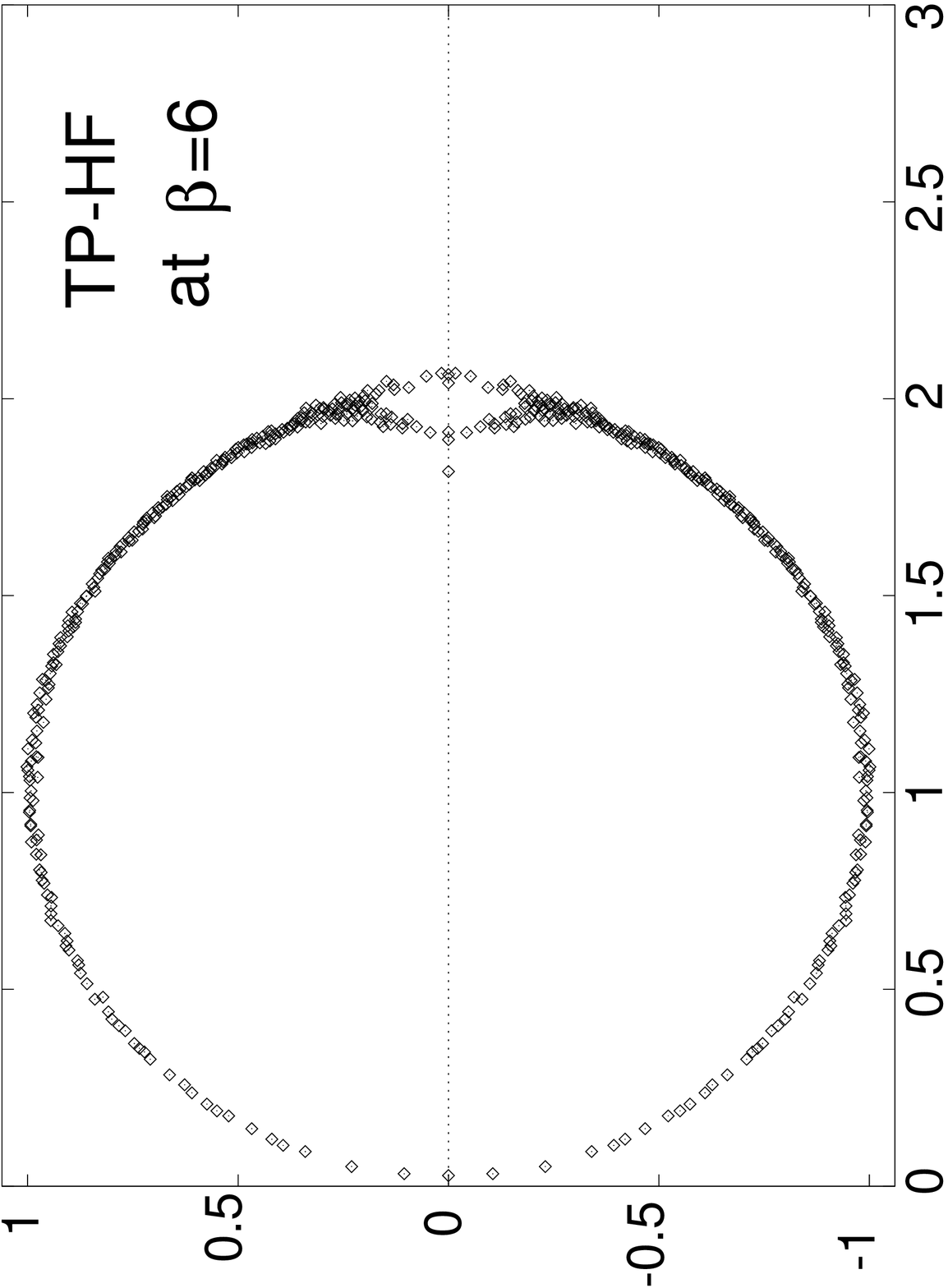} &
\hspace{-5mm}
\def\fpsangle{270} \epsfxsize=55mm \fpsbox{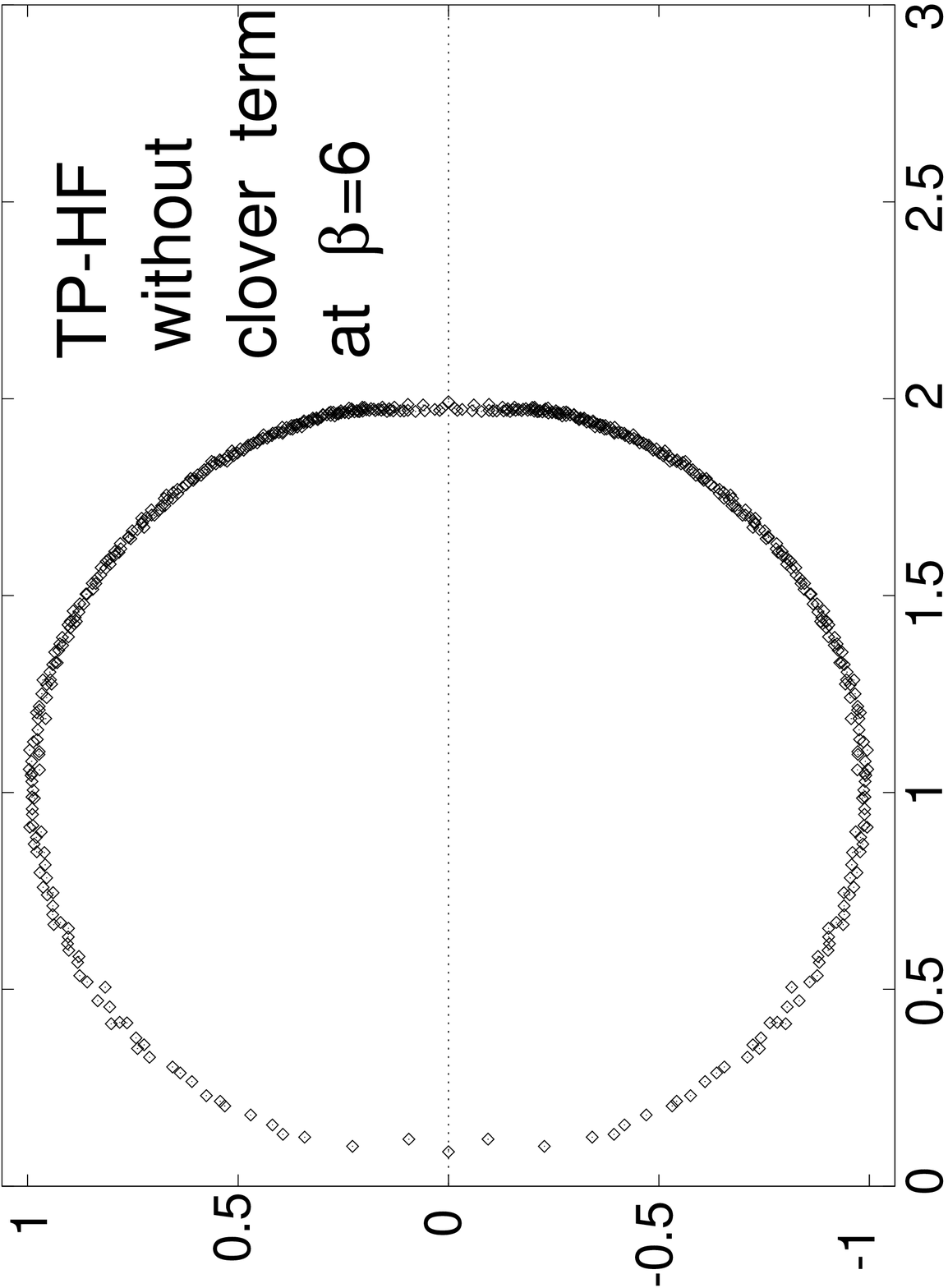} \\
\hspace{-0.6cm}
\def\fpsangle{270} \epsfxsize=55mm \fpsbox{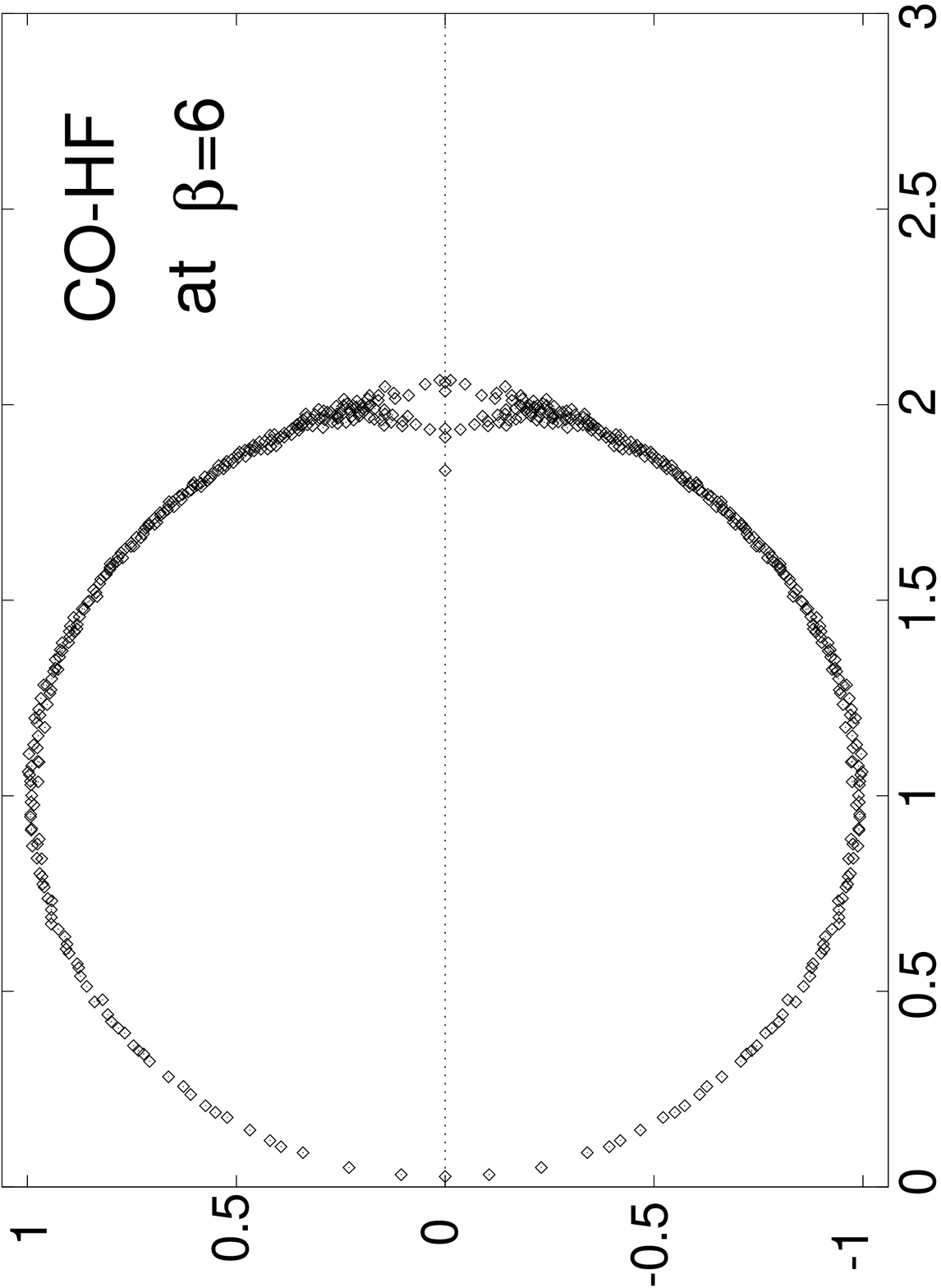} &
\hspace{-5mm}
\def\fpsangle{270} \epsfxsize=55mm \fpsbox{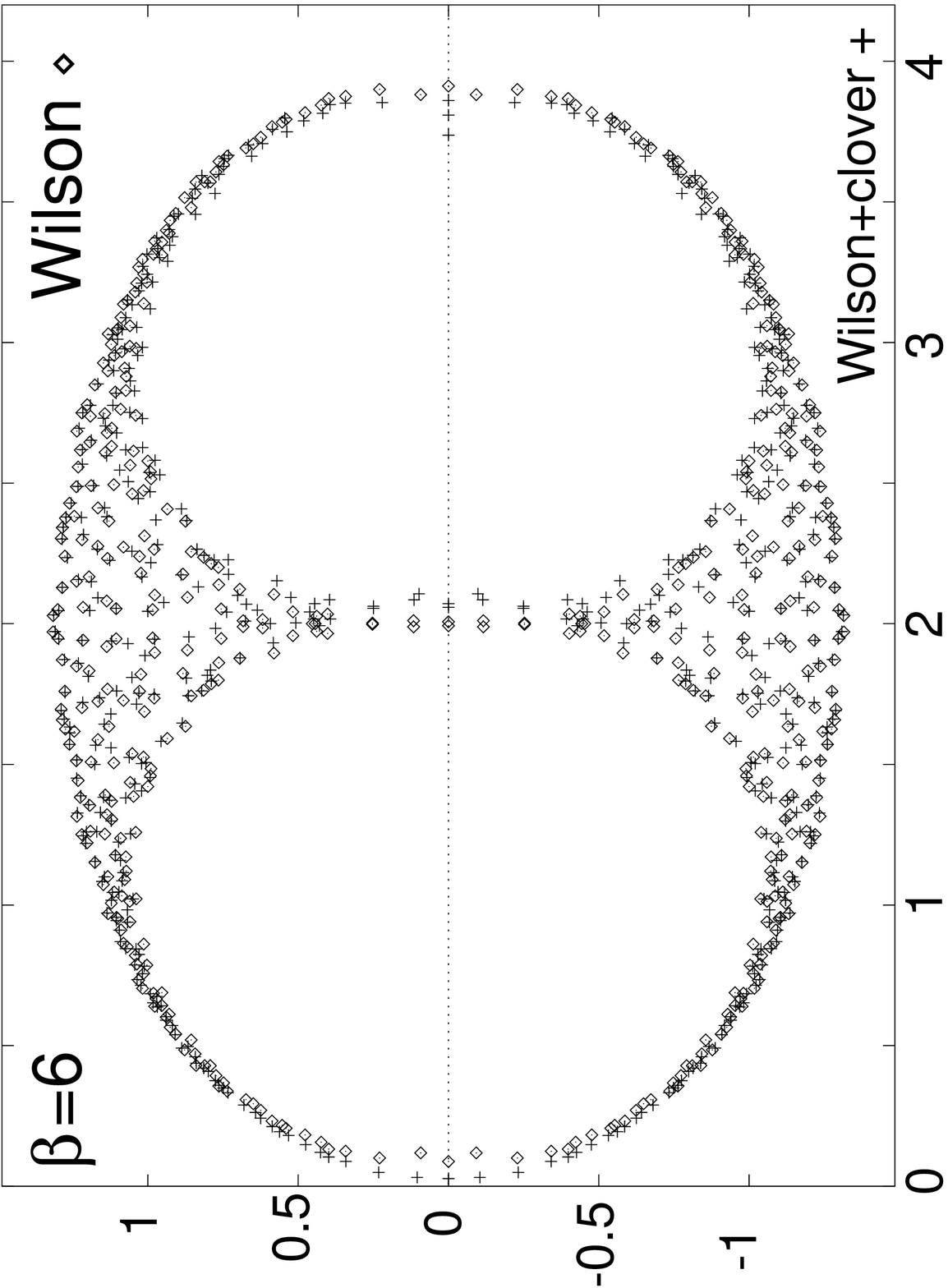}
\end{tabular}
\caption{\sl Spectra of different HFs for a typical
configuration at $\beta=6$, compared to the
Wilson fermion with and without clover term. (For the SO-HF with 
the same configuration, see Fig.\ \ref{figSO6-SO6_ov}.)
The clover term is always included in the HF actions
if not stated differently.
It has a negative impact in the region around 2,
but it improves the more important region close to 0.}
\label{figspec6}
\end{figure}

We first compare the spectra of our three HFs at $\beta =6$.
Figs.\ \ref{figspec6} and \ref{figSO6-SO6_ov}
show these spectra for a typical configuration.
They are still fairly close to the
unit circle (SO-HF is closer to a slightly larger circle, in agreement
with the observation that it has an optimal $r_{0} < 0.5$, see 
previous subsection). The splitting of the HF spectra around 2
can be viewed as a ``residue'' of the Wilson double circle.
Close to zero we observe that the smallest real eigenvalue
becomes finite. In QCD at weak coupling, such an effect
corresponds roughly to the quark mass renormalization.
Referring to this analogy, we denote the smallest real 
eigenvalues in the following as ``quark''
mass renormalization $\Delta m_{q}$. For the HFs at $\beta =6$
it amounts to $\Delta m_{q} \simeq 0.03$. 
We also see from Fig.\ \ref{figspec6} that the clover term
has actually a negative impact on the eigenvalues in the region around 2,
but it improves the more important eigenvalues close to 0.
(In this context, the Sheikholeslami-Wohlert action was studied 
in the Schwinger model in Ref.\ \cite{clover2}; for a systematic
study in $d=4$, see Ref.\ \cite{clover1}.)
In particular it decreases $\Delta m_{q}$, both, for the HFs and
for the Wilson fermion.
\footnote{Here we refer to the non-critical Wilson fermion:
$\rho^{(1)}=-\lambda_{1}=0.5$, $\rho^{(2)}=\lambda_{2}=0$.}
Interestingly the mass renormalization is practically the
same for the HFs and for the Wilson fermion with clover,
and it is again almost the same if we omit the clover term,
as it was observed before in QCD \cite{SESAM}.

\begin{figure}[hbt]
\begin{tabular}{cc}
\hspace{-0.6cm}
\def\fpsangle{270} \epsfxsize=55mm \fpsbox{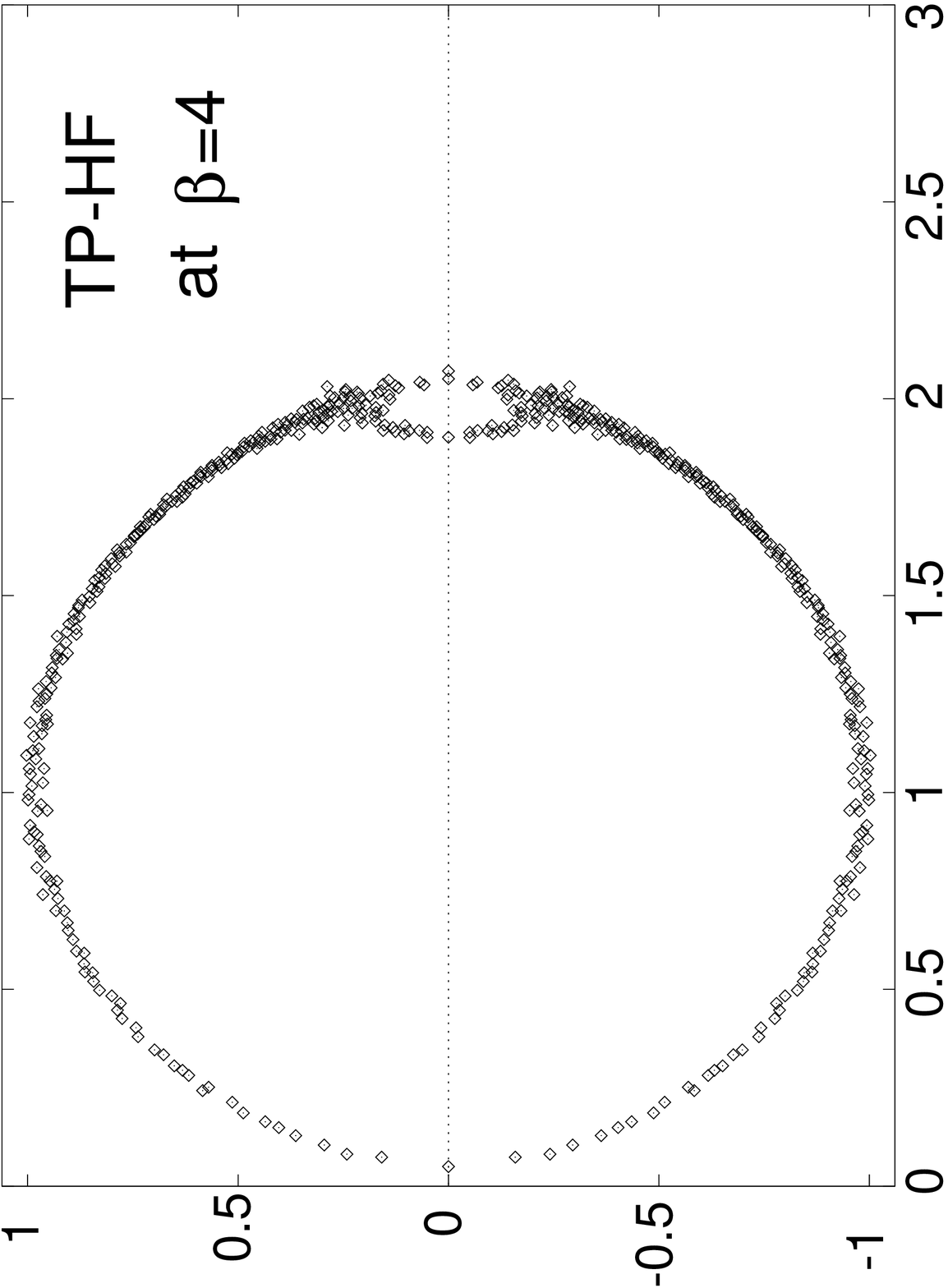} &
\hspace{-5mm}
\def\fpsangle{270} \epsfxsize=55mm \fpsbox{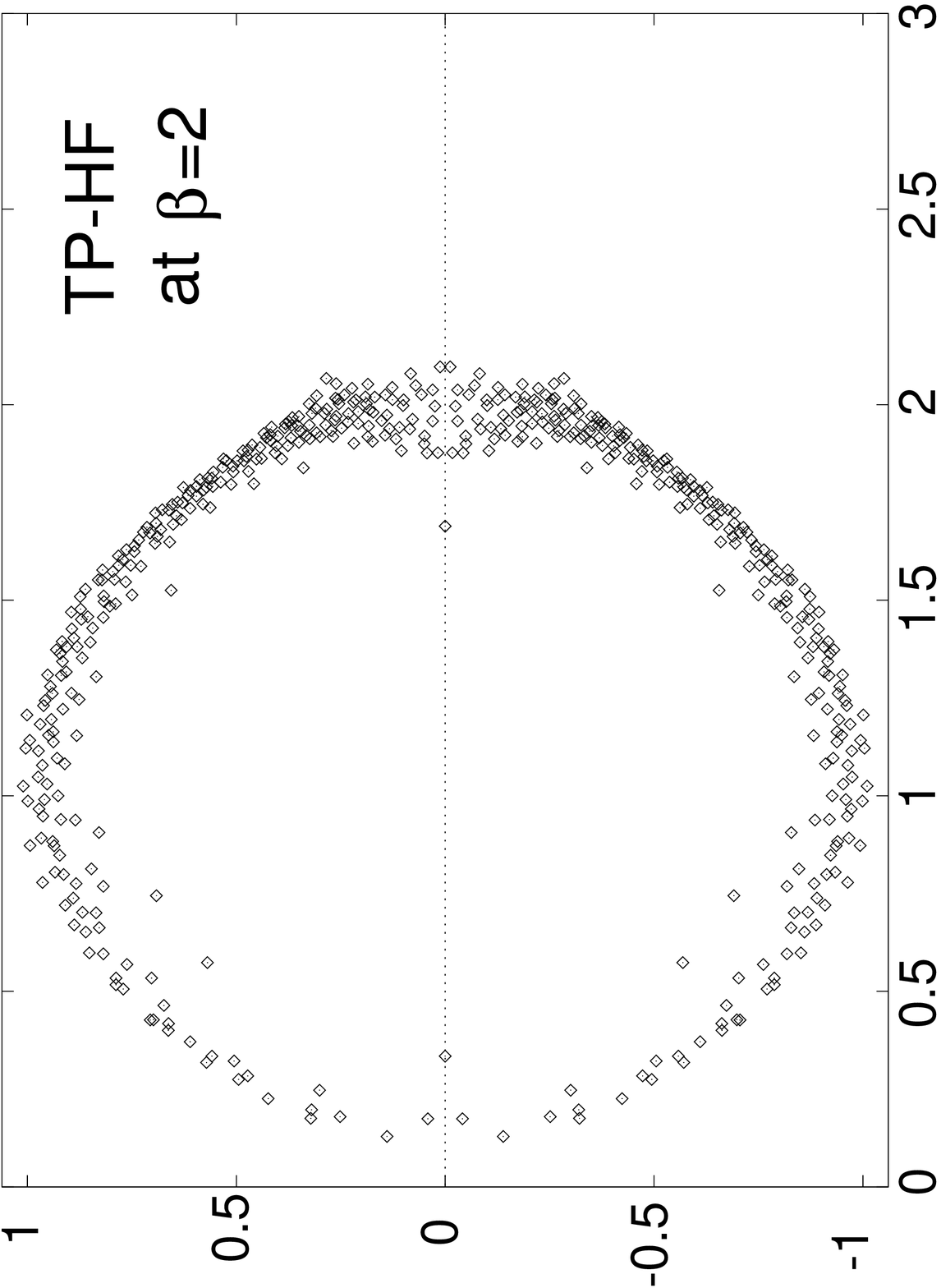}
\end{tabular}
\caption{\sl Typical spectra of the TP-HF at $\beta=4$ and $\beta=2$.
The results for the CO-HF and SO-HF are similar. Strong coupling
amplifies the deviation form the unit circle, and in particular
the mass renormalization.}
\label{figspec42}
\end{figure}

Of course, the mass renormalization --- and the deviation 
from the circle in general --- increases at stronger coupling;
as an example we show a typical TP-HF spectrum at $\beta=4$
and at $\beta =2$ in Fig.\ \ref{figspec42}. 
%(Its pattern changes only little for most other configurations.)

\begin{figure}[hbt]
%\hspace{-0.6cm}
\hspace*{25mm}
\def\fpsangle{0} \epsfxsize=80mm 
\fpsbox{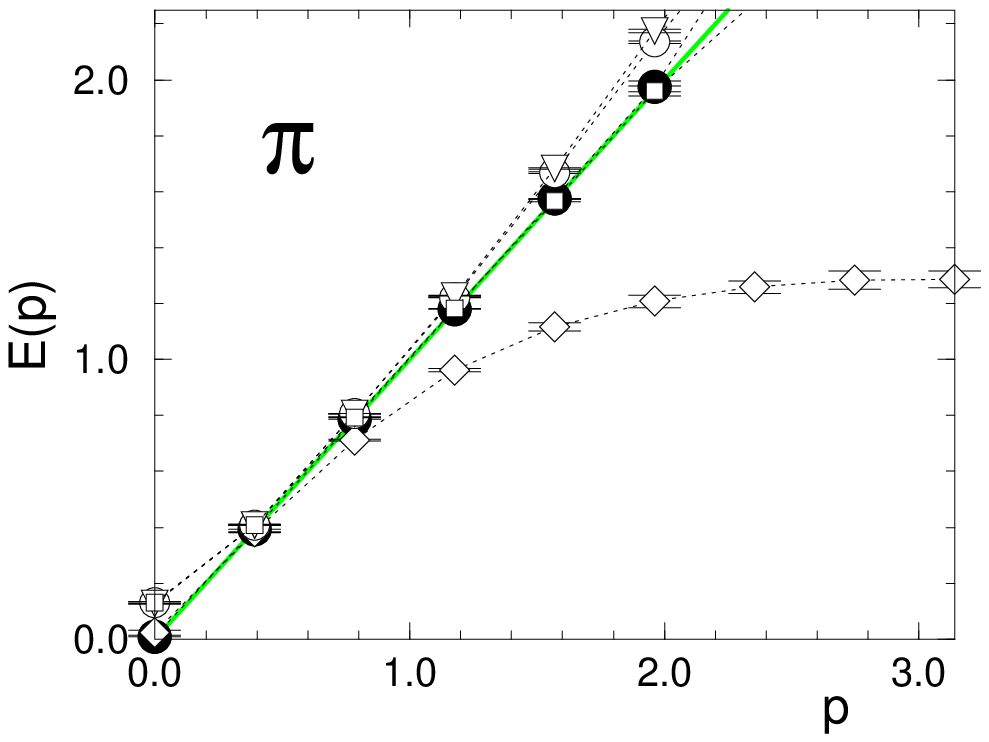}
%\end{figure}
%\begin{figure}

\hspace{25mm}
\def\fpsangle{0} \epsfxsize=80mm 
\fpsbox{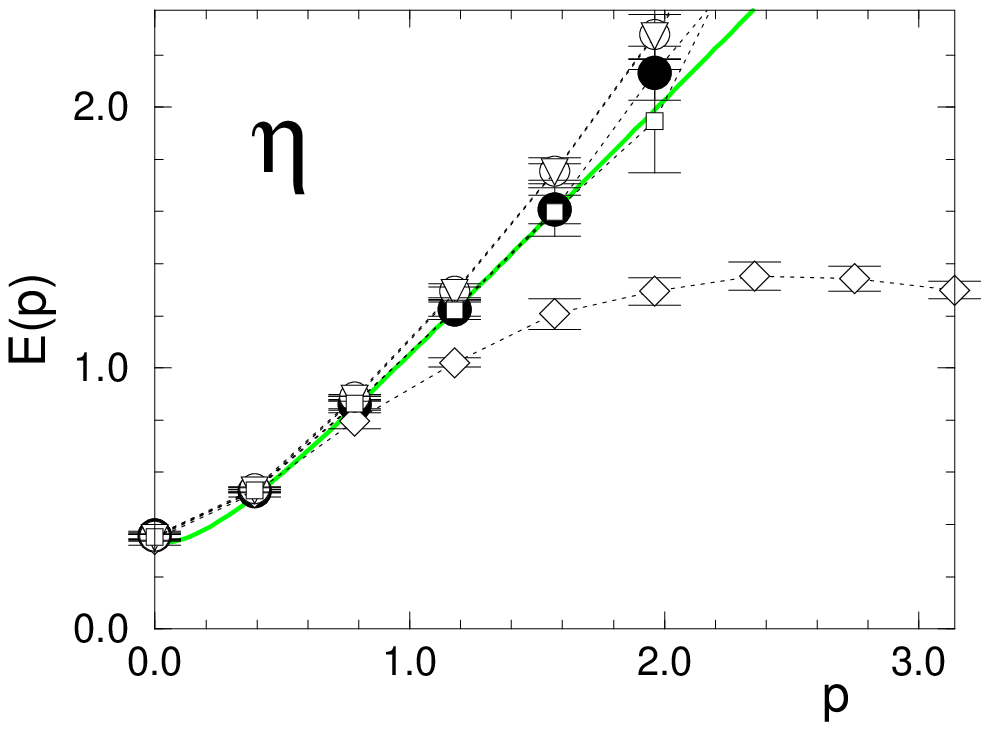}
\vspace{-5mm}
\caption{\sl Meson dispersion relations at $\beta =6$:
Wilson fermion (diamonds), TP-HF (empty circles), CO-HF
(triangles) and SO-HF (little boxes) --- all the HFs with a clover term --- 
compared to the FPA (filled circles) and the continuum (solid line).}
\label{figpi-eta6}
\vspace{-3mm}
\end{figure}

As a scaling test, we consider the dispersion relations 
of the ``mesons''; we call them ``$\pi$'' (massless) and
``$\eta$'' (massive).  In Fig.\ \ref{figpi-eta6} we show the $\pi$
and $\eta$ dispersions for our three HFs,
and we compare them to the
results for the Wilson fermion and for the FPA.
For the Wilson fermion we now use the critical
hopping parameter $\kappa_{c}= 0.25927$, which was determined
using the PCAC relation \cite{hop}.
These dispersions were obtained from 5000 configurations, using
the same ensemble for all the fermion types in Fig.\ \ref{figpi-eta6}
(as well as Fig.\ \ref{figpi-eta6_ov} in Section 3).
We see that our HFs are all strongly improved,
to the same level as the FPA. This is very remarkable,
because we only use 6 independent terms per site
--- as opposed to 123 in the FPA. 
(By contrast, the dispersions for the Sheikholeslami-Wohlert action
are very similar to the Wilson action, see first Ref.\ in \cite{clover2}.)
In addition to the excellent scaling ---
in particular for the SO-HF ---
the $\eta$ dispersion also reveals a good agreement with asymptotic 
scaling, which predicts an $\eta$ mass of 
$m_{\eta}= \sqrt{2/(\pi \beta)} \simeq 0.326$.

\begin{figure}[hbt]
%\hspace{-0.6cm}
\hspace*{17mm}
\def\fpsangle{0} \epsfxsize=100mm 
\fpsbox{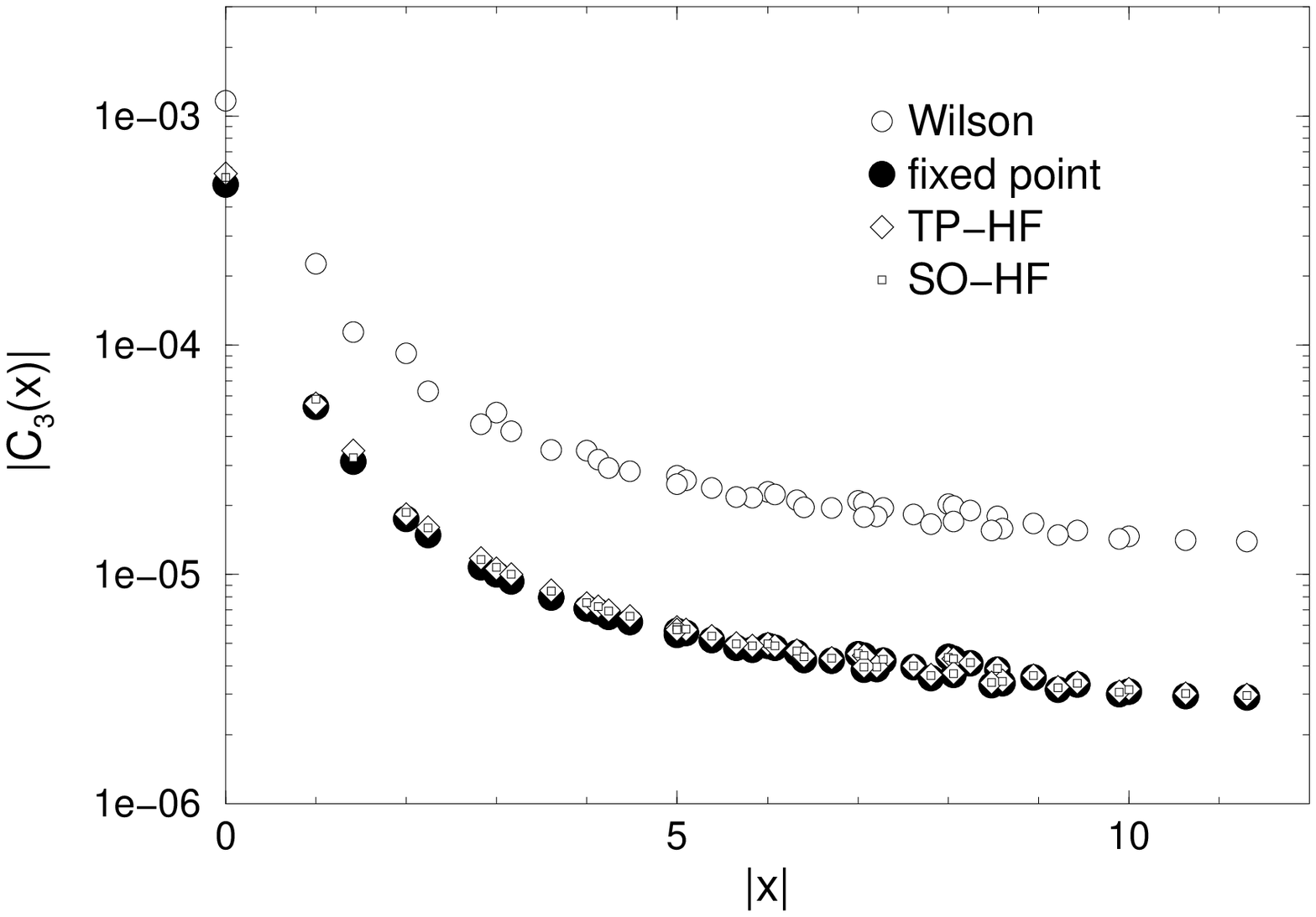}
\caption{\sl The decay of the correlation function $\vert C_{3}(x)\vert$
(defined in eq.\ \ref{C3}),
illustrating the level of approximate rotation invariance
for various fermion actions.}
\label{figrot22}
\end{figure}

As a further test for the quality of our fermion actions,
we show in Fig.\ \ref{figrot22} the decay of the correlation function
\begin{equation} \label{C3}
C_{3}(x) = \langle \bar \psi (0) \sigma_{3} \psi (0)
\cdot \bar \psi (x) \sigma_{3} \psi (x) \rangle .
\end{equation}
We are particularly interested how well rotational invariance
is approximated (which can only be measured by using 
$\sigma_{3}$ \cite{LP}). All our HFs yield a very {\em smooth}
decay --- again on the same level as the FPA ---
which reveals an excellent approximation of rotational
invariance already at short distances. 
(The jumps around $\vert x \vert \sim 8$
are finite size effects).
On the other hand, for the Wilson fermion the decay performs
a zigzag at short distances, which corresponds
to a ``taxi driver metrics''
\footnote{In the ``taxi driver metrics'', the distance
between two lattice sites is given by the shortest
lattice path(s) connecting them.};
it takes rather large distances to approximate the 
Euclidean metrics instead.

Finally we also compared the percentage of configurations,
which fulfill the index theorem, if we use the geometric
definition for the topological charge. Here the results for
the SO-HF are similar to those for the Wilson fermion
at $\beta =6$, 4 and 2, which were reported in Ref.\ \cite{FHL}.
The percentages amount to 100 \%, 99.5(2) \% and 78.7(13) \%,
respectively.\\
%%% just as the SW fermion (?) .\\

Our HFs are successful with respect to scaling and
rotational invariance, and they approximate the
GWR reasonably well up to moderate coupling strength.
Their one unpleasant feature is the quite significant mass
renormalization, which is visible again in Fig.\ \ref{figpi-eta6}.
For the $\pi$ mass it amounts to $m_{\pi} \simeq 0.13$
(which is consistent with $\Delta m_{q}$).
This is a practical problem: in this formulation,
$m_{\pi}=0$ would require the tuning of a negative bare mass.
For the massive TP-HF that is also unfavorable for locality.
However, we do know the TP-HF at any finite bare mass, and we can
construct also massive SO-HFs, see appendix. Here we do not 
simulate them --- that would also contradict our attempt to stay
close to the GWR --- but in $d=4$ this option should be
reconsidered in order to tackle the problems related
to the mass renormalization, which are more difficult there.

Indeed, the same problem was a major obstacle in similar QCD
simulations \cite{MIT,TdGHF,charm,SESAM}.
There the critical bare mass could be shifted to 0 by the use 
of fat links with {\em negative} staple terms \cite{Norbert}.
We also performed tests with fat links in the present
framework, using both, positive and negative staple terms, 
and we could remove the mass renormalization here too.
It is profitable to distinguish a staple term with weight
$w_{l} \lambda_{1}$ attached to the scalar term, and an 
independent staple weight $w_{r} \rho^{(1)}$ attached to
the vector term. Then the weight of the direct link is 
$(1 - 2 w_{l}) \lambda_{1}$ resp.  $(1 - 2 w_{l}) \rho^{(1)}$.
\footnote{An attempt to optimize the fat links analytically
--- so that the GWR violation is minimized --- yielded positive
$w_{l}$, $w_{r}$, which led, however, to even larger values
for $\Delta m_{q}$.}
In Table \ref{tabstap} we give results from various staple
combinations for the ``quark'' mass renormalization $\Delta m_{q}$
and for the mean value of the squared distance
from the unit circle, $\langle \delta^{2}_{r} \rangle$. 
The first quantity refers to the (physically essential)
small real eigenvalues, and the second quantity to the whole
spectrum. In particular, $\Delta m_{q} =0$ requires a strongly negative
$w_{l}$, which amplifies, however, $\langle \delta_{r}^{2} \rangle$.
As a remedy for that we can choose a positive $w_{r}$ (and modify
$w_{l}$ a little). However, here we optimize the GWR only,
and it turned out that ``GWR optimal'' staples are unfavorable
for the scaling behavior. The $\pi$ dispersion relation looks
still fine, but the $\eta$ dispersion is somewhat distorted.
For that reason --- and for the sake of the simplicity of the action ---
we do not use fat links in the rest of this paper, i.e.\ we return
to $w_{l}=w_{r}=0$.

\begin{table}
\begin{center}
\begin{tabular}{|c|c|c|c|}
\hline
\hline
$w_{l}$ & $w_{r}$ & $\Delta m_{q}$ & $\langle \delta_{r}^{2}\rangle$ 
(in units of $10^{-3}$) \\
\hline
\hline
   0  &     0     &    0.032(8)    &     1.34(4)   \\
\hline
%   0  &  -0.1     &    0.035    &     1.76   \\
%\hline
   0  &   0.1     &    0.029(8)    &     1.07(3)   \\
\hline
   0  &   0.2     &    0.027(7)    &     0.91(3)   \\
\hline
%   0  &   0.3     &    0.230    &     0.83   \\
%\hline
-0.1  &     0     &    0.021(6)    &     1.46(5)   \\
\hline
%-0.1  &  -0.1     &    0.024    &     1.83   \\
%\hline
-0.3  &     0     &   -0.001(4)    &     2.07(14)   \\
\hline
%-0.25 &   0.1     &    0.002    &     1.61   \\
%\hline 
-0.24 &   0.2     &    0.002(4)    &     1.59(11)   \\
\hline
\hline
\end{tabular}
\end{center}
\caption{\sl Two characteristic quantities to measure the
approximation of the standard GWR: the ``quark'' mass renormalization
$\Delta m_{q}$, and the mean value of the squared radial deviations
of the eigenvalues from the unit circle (in the complex plane),
$\langle \delta^{2}_{r} \rangle$. We show results
for varying fat links, with the scalar staple coefficient
$w_{l}$ and the vector staple coefficient $w_{r}$, at $\beta =6$.
Criticality can be achieved by a strongly negative $w_{l}$.}
\label{tabstap}
\end{table}

Instead, we are going to suggest other solutions for the
mass renormalization problem in the following
two Sections. The goal is to further reduce the GWR violation,
since $\Delta m_{q}$ and $m_{\pi}$ vanish continuously as
${\cal V}$ approaches 0. This is in agreement with the consideration
in Ref.\ \cite{Has}, where a small fermion mass is introduced
as a regularization before taking the chiral limit 
(in the sense of the GWR).

% spectrum, meson dispersion, rotation invariance, 
% (mass renormalization vanishes continuously with GWR violation;
% consistent with \cite{Has}, where a regularizing mass is used;
% major problem in QCD)
% [staples zero or further optimized]

\section{Exactly massless fermions with a good scaling behavior}

Let us introduce the operator
\begin{equation}
V = 1 - \frac{1}{\mu} D \ , \quad (\mu \neq 0)
\end{equation}
where $\mu$ is a real mass parameter,
and we require the GWR to hold with
$R_{x,y}= \frac{1}{2 \mu} \delta_{x,y}$. This is equivalent to
\begin{equation}
V^{-1} = \gamma_{5} V \gamma_{5} \ .
\end{equation}
The overlap type of solution inserts an operator $V$ of the form
\begin{equation}
V = \frac{A}{\sqrt{A^{\dagger}A}} \ .
\end{equation}
If this is well-defined, then $V$ is unitary and the GWR reads
$V^{\dagger} = \gamma_{5} V \gamma_{5}$.
Inserting some lattice Dirac operator $D_{0}$ as
\begin{equation}
A = \mu - D_{0} \ ,
\end{equation}
leads to a correctly normalized Dirac operator $D$.
Most of the literature deals with the {\em Neuberger fermion}, which 
uses $D_{0}=D_{W}$, where $D_{W}$ is the Wilson-Dirac operator.
Neuberger fermions were simulated (quenched) in the Schwinger
model \cite{FHL,SC2,FHLW} and in 4d non-Abelian gauge theory 
\cite{SCRI,HJL}. The mass parameter $\mu$ is usually set to 1.
As long as we stay with the Wilson-Dirac operator
it can be absorbed in the mass $M$ of $D_{W}$.
But if we insert a more sophisticated $D_{0}$, then $\mu$
takes a non-trivial r\^{o}le, see below.
%(If $M$ is the mass in $D_{W}$, then $\mu - M \in (0,2)$ is required
%to obtain one flavor.
%\footnote{The situation {\em on} the boundaries is 
%tricky.})
%However, one can also substitute $D_{W}$ by more sophisticated 
%lattice Dirac operators \cite{WB}, and then the mass parameter
%$\mu$ takes a non-trivial r\^{o}le.

%The locality of the Neuberger fermion has also been studied
%in QCD \cite{HJL}. Locality was established analytically for smooth
%gauge fields. At $\mu - M=1$ the dimensionally generalized
%condition is that any plaquette variable $P$ has to obey
%\begin{equation}
%\vert 1 - P \vert < \frac{2}{5d (d-1)} \ ,
%\end{equation}
%(since this rules out the danger of $A^{\dagger}A = 0$)
%and we see that this condition is more power in $d=2$
%than in $d=4$. In numerical tests down to $\beta =6$
%it turned out the the degree of locality is only reduced 
%gradually by the gauge interaction \cite{HJL}.
%The observation that it remains rather close to the free
%locality raises hope that the improvement of free locality
%persists also in gauge theory. 
It was observed that the use of a suitable $D_{0}=D_{HF}$ instead of
$D_{W}$ improves the locality of the free $D$ 
significantly \cite{WB}, and it looks promising also with other 
respects. Further properties of improved overlap fermions
will be discussed in this Section.

% General construction; improvement of free locality \cite{WB}
% Ref.\ \cite{HJL} : degree of locality decreases only
% gradually as gauge interactions are switched on.
% Hence an improved locality in the free case provides hope
% for superiority also in the interacting case.

\subsection{Free overlap fermions}

If we insert the ansatz $D_{0} = \rho_{\mu} \gamma_{\mu} + \lambda$,
then the free overlap Dirac operator and its inverse
read (in momentum space)
\begin{eqnarray} \hspace{-8mm}
D(p) &=& 1 + \frac{\rho_{\mu}(p) \gamma_{\mu} + \lambda (p) - \mu}
{\sqrt{-\rho^{2}(p) + [ \lambda (p) - \mu ]^{2} } } \ ,
\nonumber \\ \hspace{-8mm}
D^{-1}(p) &=& \frac{1}{2} \Big( 1 - \frac{\rho_{\mu}(p)\gamma_{\mu}}
{Q(p)+u(p)} \Big) , \quad Q(p) := \sqrt{-\rho^{2}(p)+u^{2}(p)}, \
u(p) := \lambda (p) - \mu \ . \label{Dp}
\end{eqnarray}
There are two conditions for a pole in the free propagator:
\begin{equation} \label{cond12}
(1) \qquad \rho^{2}(p) = 0 \ , \qquad \qquad 
(2) \qquad u(p) < 0 \ .
\end{equation}
Hence the vector term $\rho_{\mu}$ determines the shape of the
fermion dispersion, and the scalar term $\lambda$ can fix an end-point
of that dispersion curve
\footnote{It is the square root which causes this unusual behavior.}.
(This is different from the usual case,
where both terms contribute to the
shape, and the curve cannot just end inside the Brillouin zone.)
From the symmetry properties that we assumed for the vector term,
condition (1) allows for poles whenever all momentum 
components obey $p_{\mu} \in \{ 0, \pi \}$, hence it allows for
fermion doubling with $2^{d}$ species. However, the condition (2)
can still save us from doubling; $\lambda$ and $\mu$ should be
chosen such that only the pole at $p=0$ really occurs.
If we insert $D_{W}$ with a mass $M$, 
then the condition for one free species is simply
$0 < (\mu -M) < 2$.
\footnote{The situation {\em on} the boundaries is tricky.}
In gauge theory, some tuning of $\mu$ might be required to stay with
one species \cite{SC}.

\begin{figure}[hbt]
%\begin{center}
\hspace*{25mm}
\def\fpsangle{270}
\epsfxsize=70mm
\fpsbox{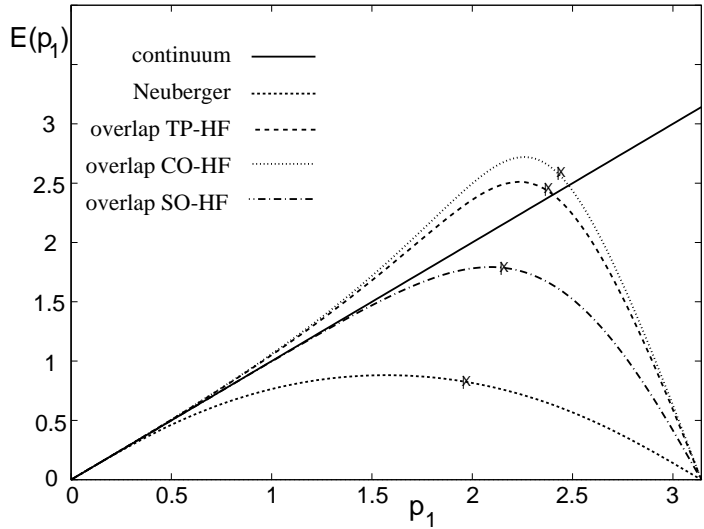}
%\end{center}
\vspace{-3mm}
\caption{\sl The free dispersion relation for the Neuberger
fermion (inserting $D_{0}=D_{W}$ in the overlap formula)
and for various improved overlap fermion (inserting
$D_{0}=D_{HF}$). We mark the end-points of the curves for 
the mass parameter $\mu =1$.}
\label{figGWfreedisp}
\end{figure}

For a HF with an even scalar term $\lambda$, the conditions read
\footnote{At this point, we do not insist on $\sum_{r} \lambda (r)=0$,
i.e.\ $D_{HF}$ may also be massive.}
\begin{equation}
\mu - \lambda_{0} - 4 ( \lambda_{1}+\lambda _{2} ) > 0 \ , \quad
\mu - \lambda_{0} + 4 \lambda_{2} < 0 \ , \quad
\mu - \lambda_{0} - 8 \lambda_{2} < 0 \ ,
\end{equation}
where we use the notation introduced in Table \ref{tabHF}.
In the presence of gauge interactions, the 1-species region
becomes more narrow from both sides. This can be understood from a
crude consideration, which introduces ``effective links''
somewhat smaller than 1, and which therefore moves the
(negative) couplings $\lambda_{1} \simeq 2 \lambda_{2}$
to larger ``effective'' values. This was also observed
numerically for the TP-HF in the Schwinger model \cite{SCpri}.
\footnote{As an example, the massless TP-HF has the free 
condition is $0 < \mu < 1.96$.
At $\beta = 3$ the physical interval shrinks to about
$\mu \in (0.3,1.8)$; outside this interval the index theorem
is violated.}
Hence we should start far from the boundaries in the free
case, to be on more solid grounds in gauge theory.
The truncation by condition (2) should occur not too close
to the edge of the Brillouin zone, but of course quite far 
from $p=0$, because this region is really needed.
It appears that $\mu= 1$ is a very reasonable choice.
In Fig.\ \ref{figGWfreedisp} we show the free fermion 
dispersion for overlap fermion $D$ constructed
from $D_{W}$ and from our three variants of $D_{HF}$
(all of them massless): we show the full curve given
by condition (1), and we mark the end-points for $\mu =1$.
Those end-points can be shifted arbitrarily by changing $\mu$.
(Higher branches do not appear, see appendix.)

We see that the overlap dispersion based on $D_{W}$
is unfortunately worse than the dispersion of the ordinary
Wilson fermion, cf.\ Fig. \ref{figfreedisp}. 
(Technically, the reason is that
in the ordinary case the scalar term helps to raise the curve
a little). It has been conjectured that overlap fermions
might be free of $O(a)$ artifacts even in gauge theory.
However, for the Neuberger fermion the $O(a^{2})$ artifacts 
are quite bad, as we see even in the free case.
The improvement by using $D_{HF}$ is again very significant,
in particular for the  SO-HF.

\begin{figure}[hbt]
\begin{tabular}{cc}
\hspace{-0.6cm}
\def\fpsangle{270} \epsfxsize=55mm \fpsbox{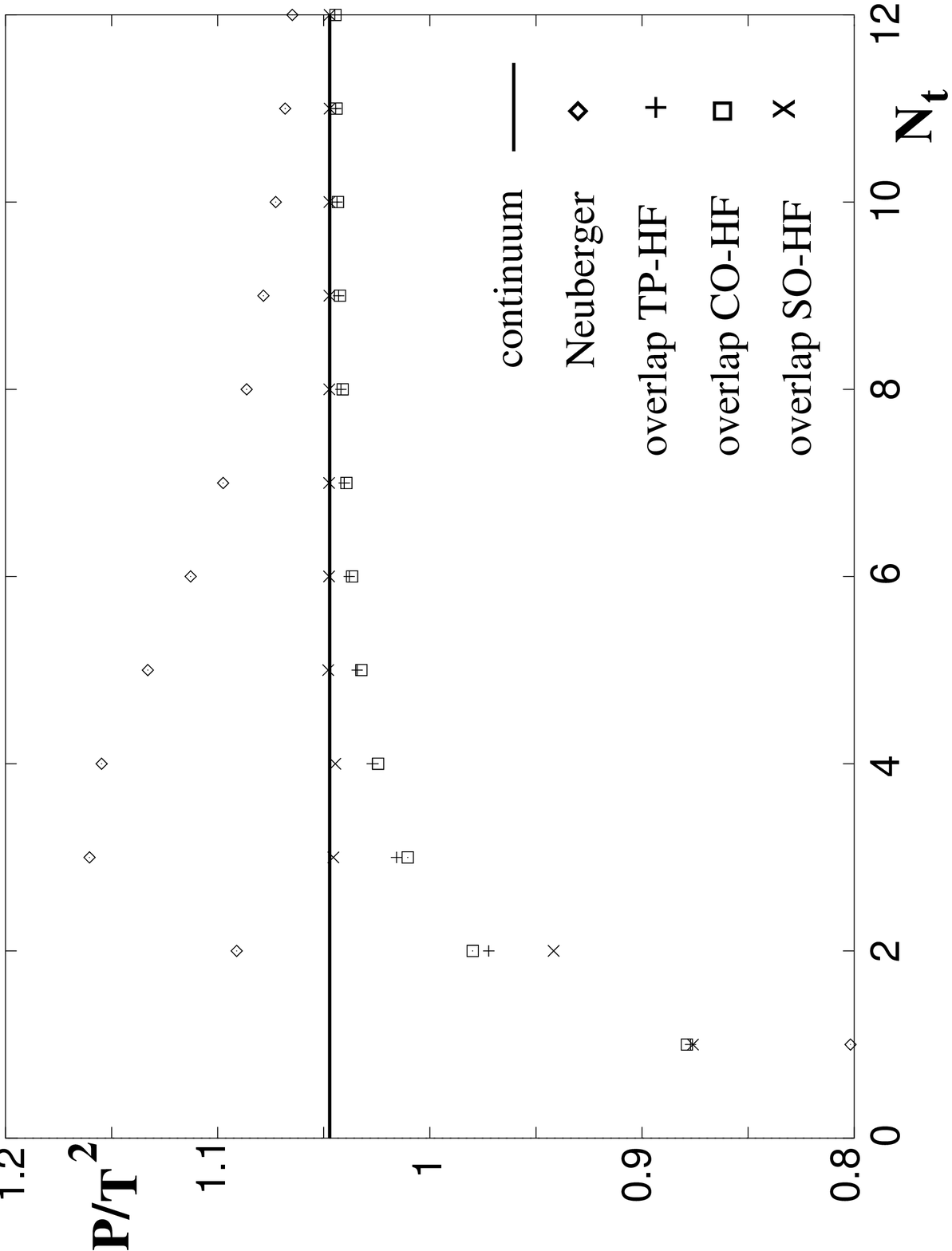} &
\hspace{-5mm}
\def\fpsangle{270} \epsfxsize=55mm \fpsbox{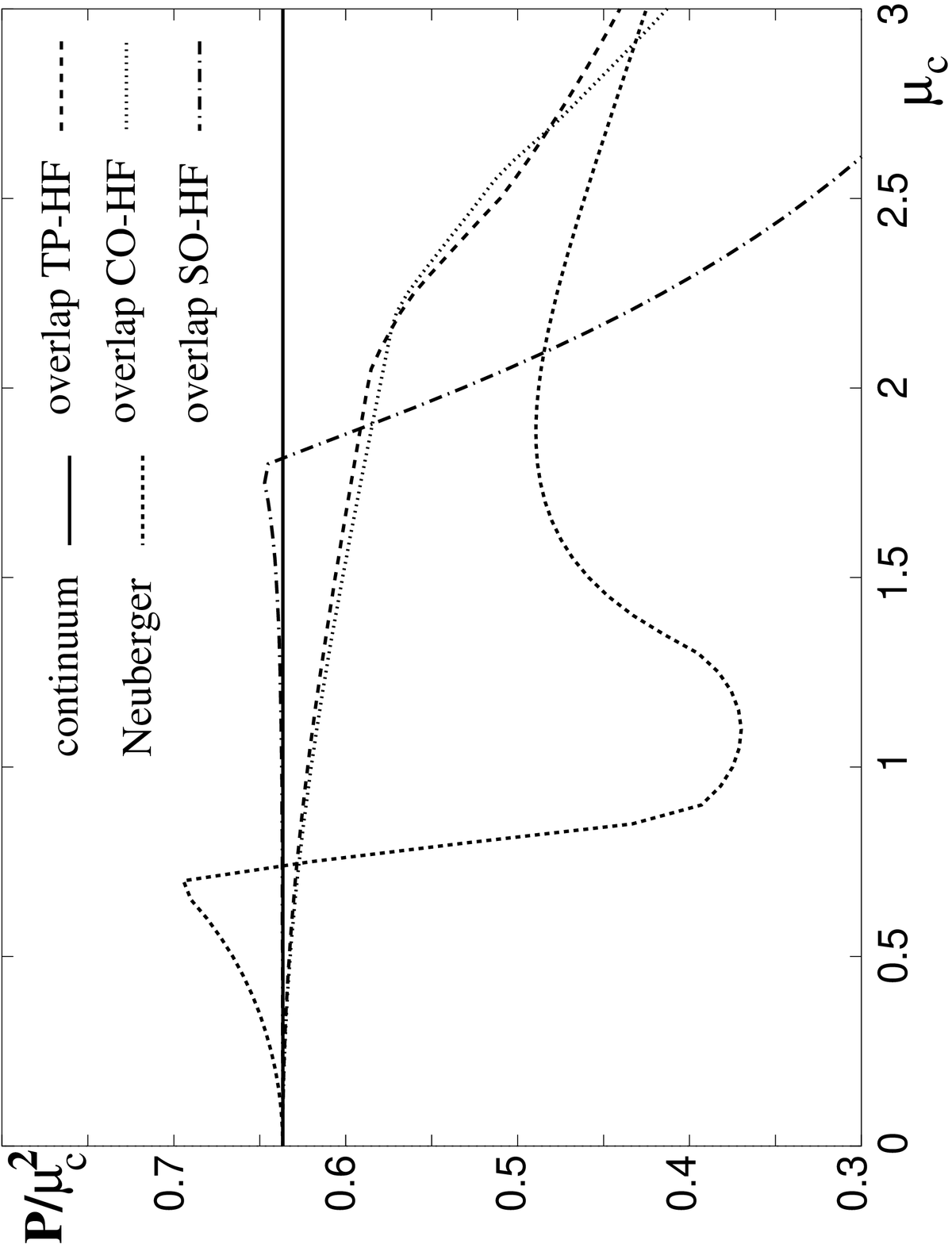}
\end{tabular}
\caption{\sl Thermodynamic scaling ratios:
$P/T^{2}$ at $\mu_{c}=0$ (left), and $P/\mu_{c}^{2}$ at $T=0$ (right),
as in Fig.\ \ref{figthermo}, but now for the corresponding
overlap fermions. Again the overlap SO-HF is clearly most successful.}
\label{figthermo_ov}
\end{figure}

The improved scaling of free overlap-HFs is also confirmed if
we repeat our thermodynamic considerations, see Fig.\ \ref{figthermo_ov}.
The good quality of these fermions does hardly come as a
surprise \cite{WB}: if we insert a fermion into the overlap formula,
which obeys the GWR for $R = \delta_{x,y}/(2\mu )$, then
it reproduces itself due to $A^{\dagger} A = \mu^{2}$. Now we insert
an approximate GW fermion, so its modification by the overlap
formula is rather modest, and the good scaling quality does
essentially persist.
\footnote{Here the overlap correction $A \to \mu A /\sqrt{ 
A^{\dagger} A}$ is completely obvious.}

In the appendix we comment on the use of massive $D_{HF}$ ---
which still produce massless overlap fermions --- and also
on the option to render overlap fermions massive themselves.

%fermion dispersion, thermodynamics \\
%start from good HF (chiral + scaling) $\to$ minor modification,
%good scaling and rotation invariance cannot be completely destroyed

\subsection{Interacting overlap fermions}

\begin{figure}[hbt]
%\hspace{-0.6cm}
\hspace*{25mm}
\def\fpsangle{270} \epsfxsize=70mm 
\fpsbox{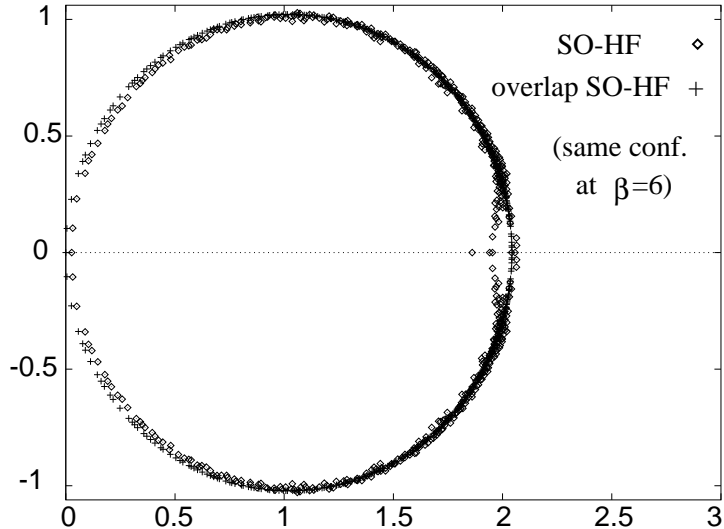}
\caption{\sl The mapping of the eigenvalues of the SO-HF
(for a configuration at $\beta=6$) onto the unit circle
(in the complex plane) by means of the overlap formula.}
\label{figSO6-SO6_ov}
\end{figure}

Also in the interacting case, the overlap-Dirac operator
\begin{equation}
D = \mu \, \Big[ \, 1 - \frac{A}{\sqrt{A^{\dagger}A}} \, \Big]
\end{equation}
satisfies the GWR with $R_{x,y} = \delta_{x,y} / (2 \mu )$,
hence its spectrum is situated on the circle with center
and radius $1/\mu $.
We now wonder what really happens to the eigenvalues
of $D_{HF}$ as the hypercube fermion is inserted in the
overlap formula. Fig.\ \ref{figSO6-SO6_ov} shows as an example those
eigenvalues (for a typical configuration
at $\beta =6$ on our $16\times 16$ lattice)
before and after application of the overlap formula.
Especially in the region around zero --- where the eigenvalue
density is low --- there is an obvious mapping of the eigenvalues
one by one onto the circle. In such cases, the effect of the overlap
formula is close to a radial projection of each eigenvalue
onto the circle. This can easily be understood from the overlap
formula, and it is in agreement with the above statement
that the overlap modifies an approximate GW fermion only modestly.
(Of course this observation is not so obvious if we insert
a fermion far from a GW fermion, such as $D_{W}$.)

This geometric understanding of the overlap formula
also provides a neat interpretation of the parameter $\mu$:
it is simply the center of the circle (through 0) that the spectrum
is projected on. Of course that center must be chosen between
the small real eigenvalues of the original spectrum (which
have to be mapped to 0) and the larger real eigenvalues
(to be mapped on $2\mu$). So the allowed range of $\mu$
can be recognized immediately from the spectrum of $D_{0}$, see
e.g.\ Figs.\ \ref{figspectrumHF}, \ref{figspec6}, \ref{figspec42}
and \ref{figSO6-SO6_ov}. This range shrinks at stronger
coupling. If we do not respect it, then we map one or several
eigenvalues to the wrong arc crossing the real axis,
and as a consequence the index theorem is violated
(as we mentioned already in the previous subsection).

\begin{figure}[hbt]
%\hspace{-0.6cm}
\hspace*{25mm}
\def\fpsangle{0} \epsfxsize=80mm 
\fpsbox{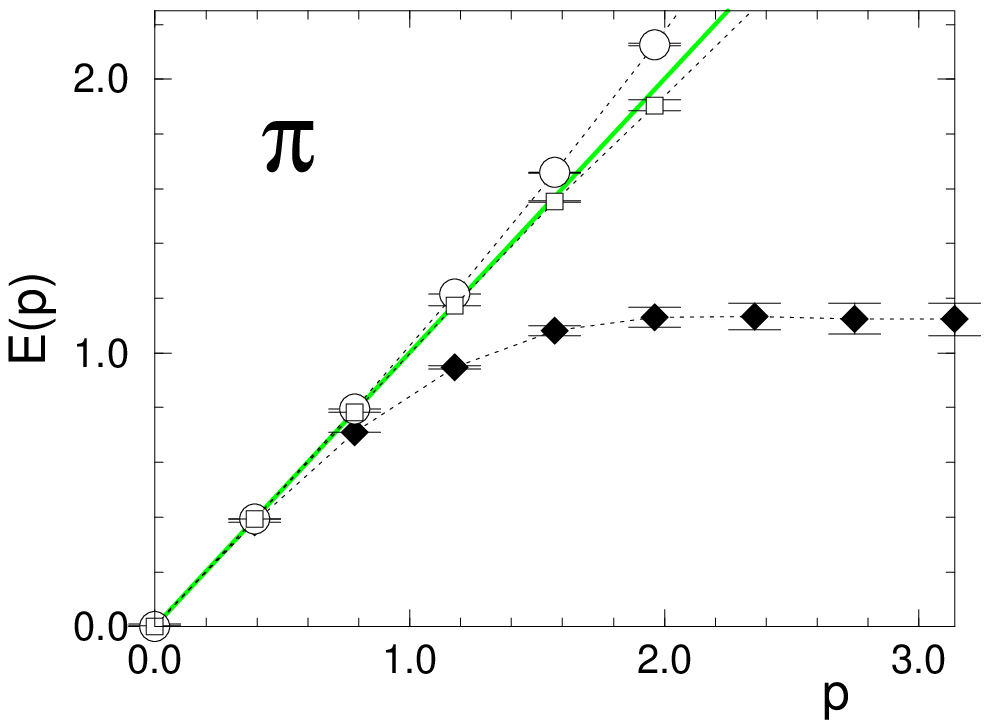}
%\end{figure}
%\begin{figure}

\hspace{25mm}
\def\fpsangle{0} \epsfxsize=80mm 
\fpsbox{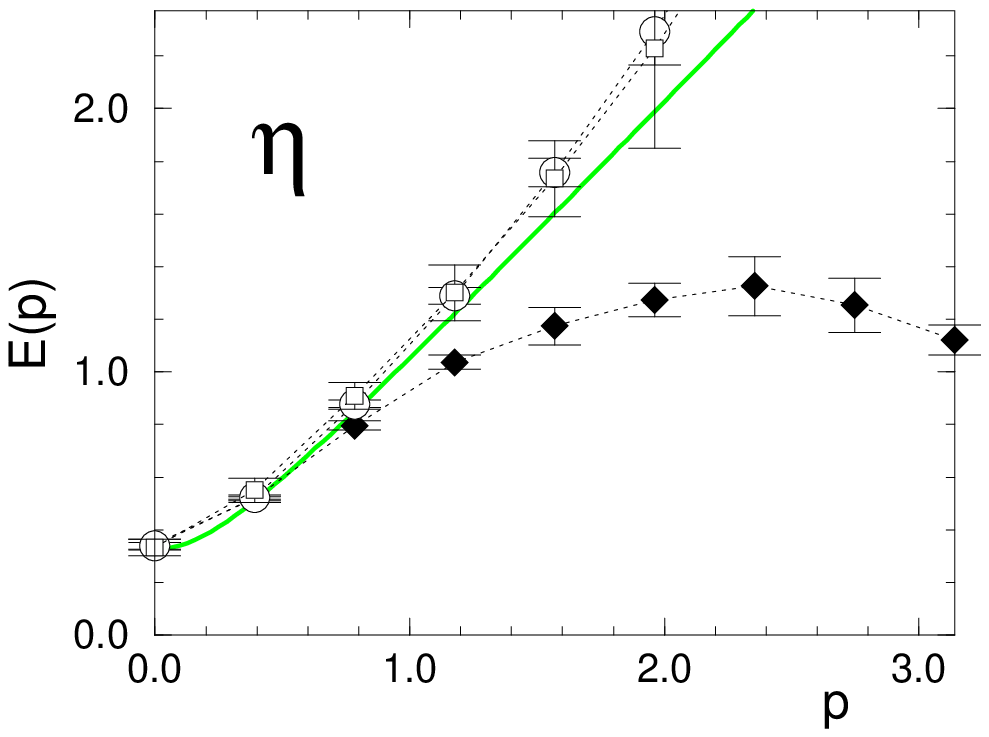}
%\vspace{-5mm}
\caption{\sl Meson dispersion relations at $\beta =6$:
Neuberger fermion (diamonds), overlap TP-HF (open circles)
and  the overlap SO-HF (little boxes) compared to the continuum
(solid line).}
\label{figpi-eta6_ov}
\vspace{-3mm}
\end{figure}

\begin{figure}[hbt]
%\hspace{-0.6cm}
\hspace*{17mm}
\def\fpsangle{0} \epsfxsize=100mm
\fpsbox{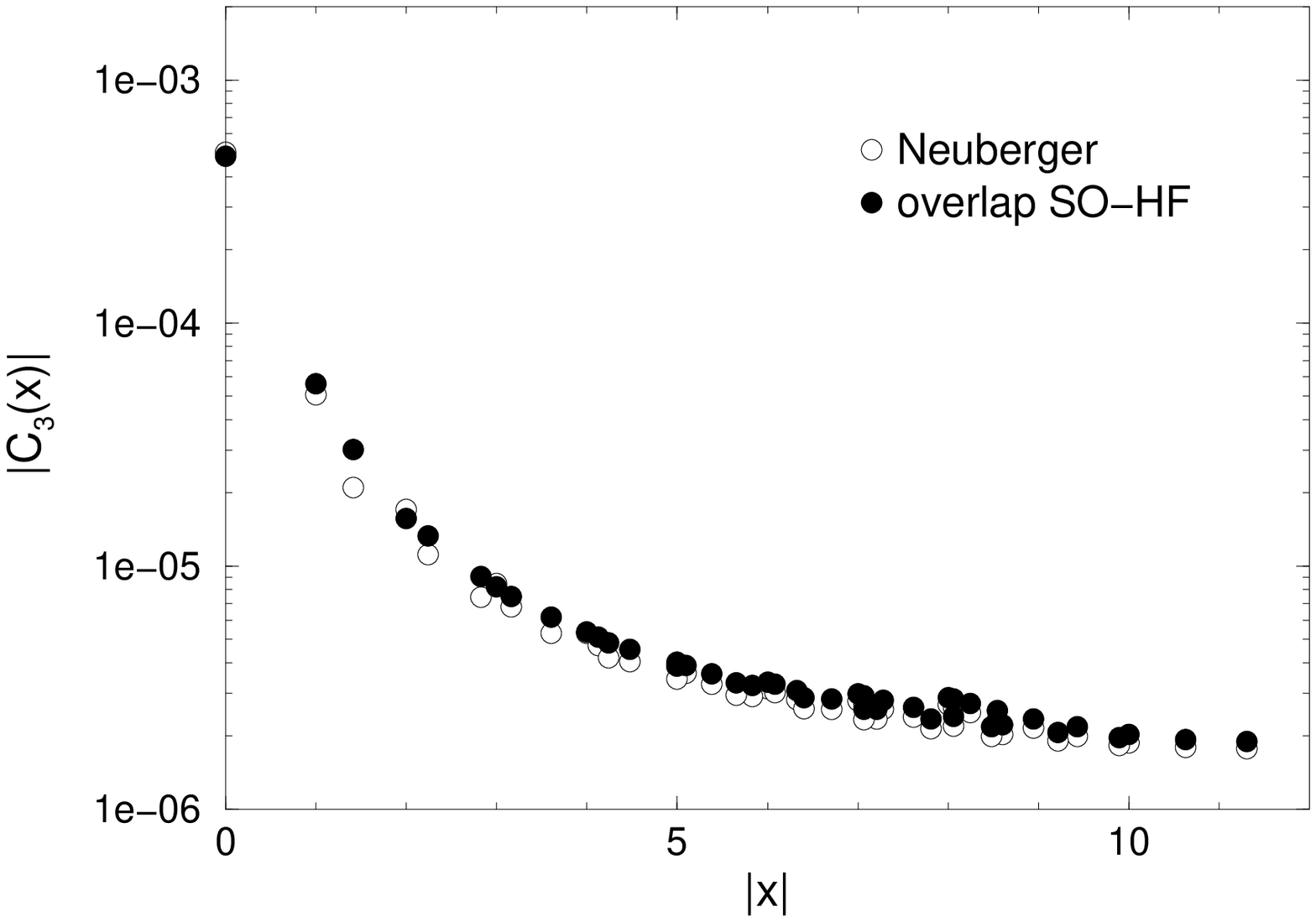}
\vspace{-3mm}
\caption{\sl The decay of the correlation function $\vert C_{3}(x)\vert$
(defined in eq.\ \ref{C3}),
illustrating the level of approximate rotation invariance
for two overlap fermions.}
\label{figrot32}
\end{figure}

These observations raise hope that also in gauge theory the scaling
of an approximate GW fermion is not much affected
by the ``{\em chiral projection}''. Indeed, Figs. \ref{figpi-eta6_ov} 
and \ref{figrot32}
show that the scaling quality --- tested by the meson dispersion ---
as well as the approximate rotational invariance is still very good,
whereas the Neuberger fermion is contaminated by considerable
artifacts (again it is a little worse than the ordinary Wilson fermion).
In particular for the overlap SO-HF the $\pi$ dispersion is still 
excellent, but the $\eta$ dispersion is a little distorted by the 
overlap formula (the slope is a bit too steep).
The same effect occurs if we remove the mass renormalization
by means fat links (using a negative $w_{l}$, cf. Section 2.2).\\

As a further issue, we now want to address the question of locality.
For the Neuberger fermion in $d=4$, locality was established 
analytically for very smooth gauge fields \cite{HJL}. 
At $\mu - M=1$ the dimensionally generalized
condition is that any plaquette variable $P$ has to obey
%(the norm is given by the largest absolute value of an eigenvalue)
\begin{equation}
\Vert 1 - P \Vert < \frac{2}{5d (d-1)} \ ,
\end{equation}
since this rules out the danger of $A^{\dagger}A = 0$.
We see that this condition is more powerful in $d=2$
than in $d=4$. 

\begin{figure}[hbt]
\begin{tabular}{cc}
\hspace{-0.5cm}
\def\fpsangle{270} \epsfxsize=55mm \fpsbox{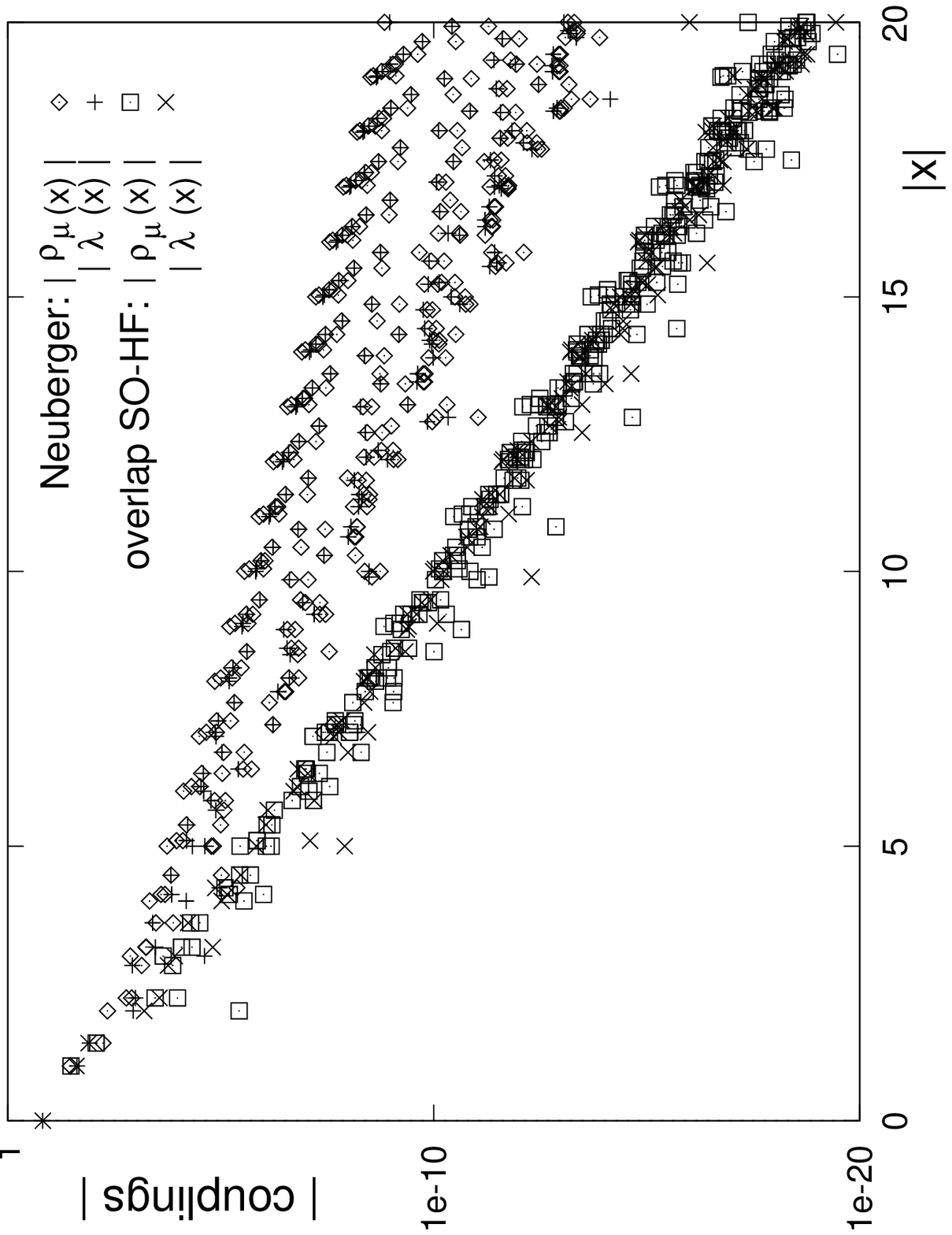} &
\hspace{-6mm}
\def\fpsangle{0} \epsfxsize=77mm \fpsbox{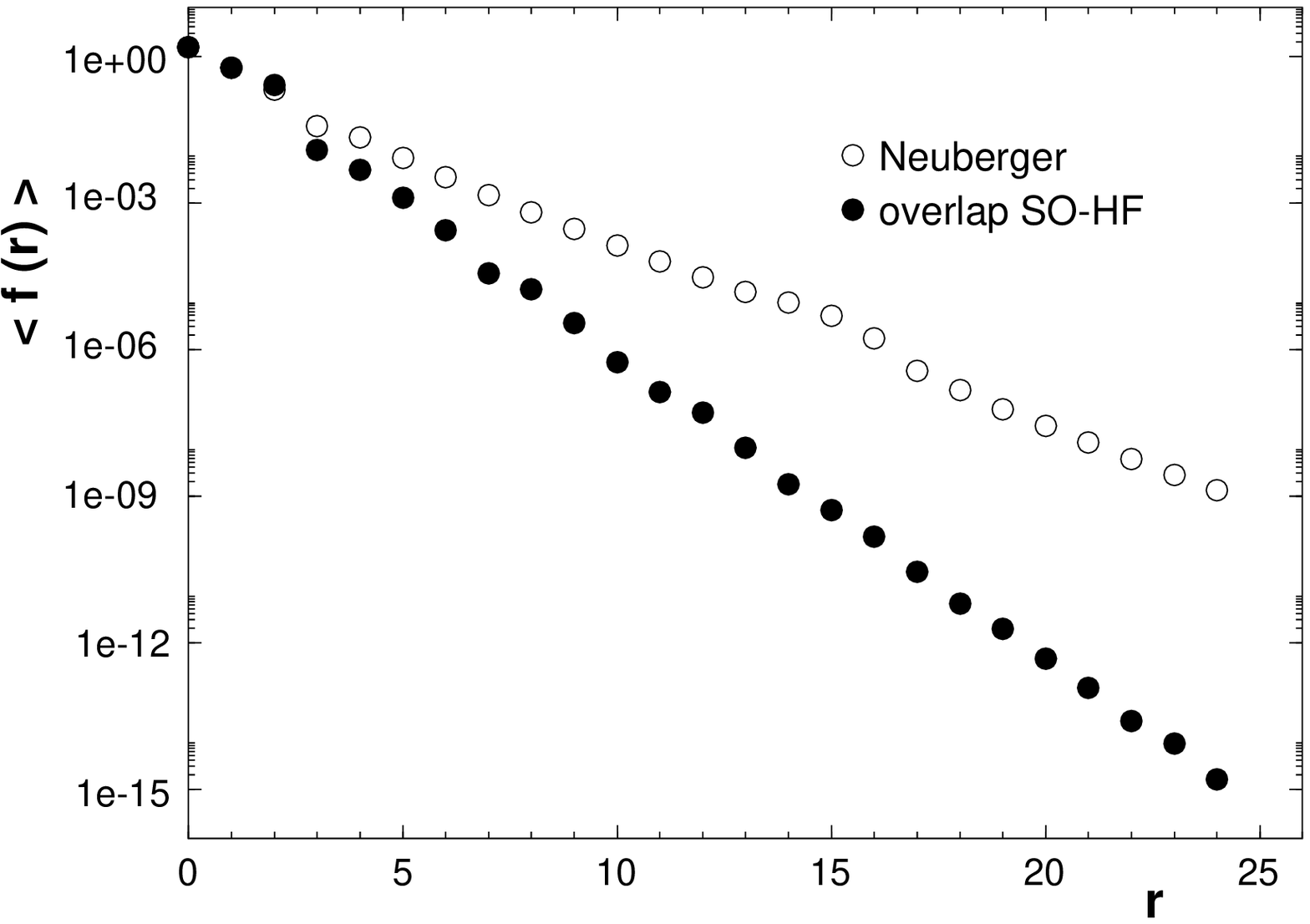}
\end{tabular}
\caption{\sl The locality of the overlap SO-HF compared to the
Neuberger fermion. Left: the decay of the free couplings
(as introduced in eq.\ (\ref{freefermi})).
Right: the decay of the maximal correlation
(as defined in eq.\ (\ref{ff})) at $\beta = 6$ on a 
$24\times 24$ lattice.}
\label{figloc}
\end{figure}

In numerical tests in QCD down to $\beta =6$,
it turned out that the degree of locality is only reduced 
gradually by the gauge interaction \cite{HJL}.
\footnote{As the interaction is turned on, the value of
$\mu -M$, which is optimal for locality, moves somewhat above $1$.}
The observation that it remains rather close to the free
locality raises hope that the improvement of free locality
persists also in gauge theory. 
In fact, this is confirmed in our Schwinger model study
as we see from Fig.\ \ref{figloc}. It compares
first the free locality, and then the absolute value
of the maximal correlation over a distance $\vert x \vert $
at $\beta =6$ on a $24 \times 24$ lattice.
More precisely, we show the expectation value of
\begin{equation} \label{ff}
{\bf f}({\bf r}) = \ ^{\rm max}_{~~y} \ \{ \ \Vert \psi (y) \Vert 
\ \rule[-1.3mm]{0.4mm}{5mm}  \ \vert x-y \vert = {\bf r} \ \}
\end{equation}
for a unit source at $x$, as it was done before for the Neuberger
fermion in QCD \cite{HJL}.
We see that the exponential decay is much faster for the
overlap SO-HF than it is the case for the Neuberger fermion.
(Again, the little bumps in the middle of that plot are finite size effects;
they move continuously if we vary the lattice size.)
%\begin{figure}[hbt]
%%\hspace{-0.6cm}
%\hspace*{25mm}
%\def\fpsangle{0} \epsfxsize=80mm 
%\fpsbox{loc.eps}
%\vspace{-3mm}
%\caption{\sl The locality of the overlap SO-HF (circles) compared to the
%Neuberger fermion (diamonds): we show the decay of the maximal correlation
%(as defined in eq.\ (\ref{ff})) at $\beta = 6$.}
%\label{figloc}
%\end{figure}

\begin{figure}[hbt]
\begin{tabular}{cc}
\hspace{-0.5cm}
\def\fpsangle{0} \epsfxsize=76mm \fpsbox{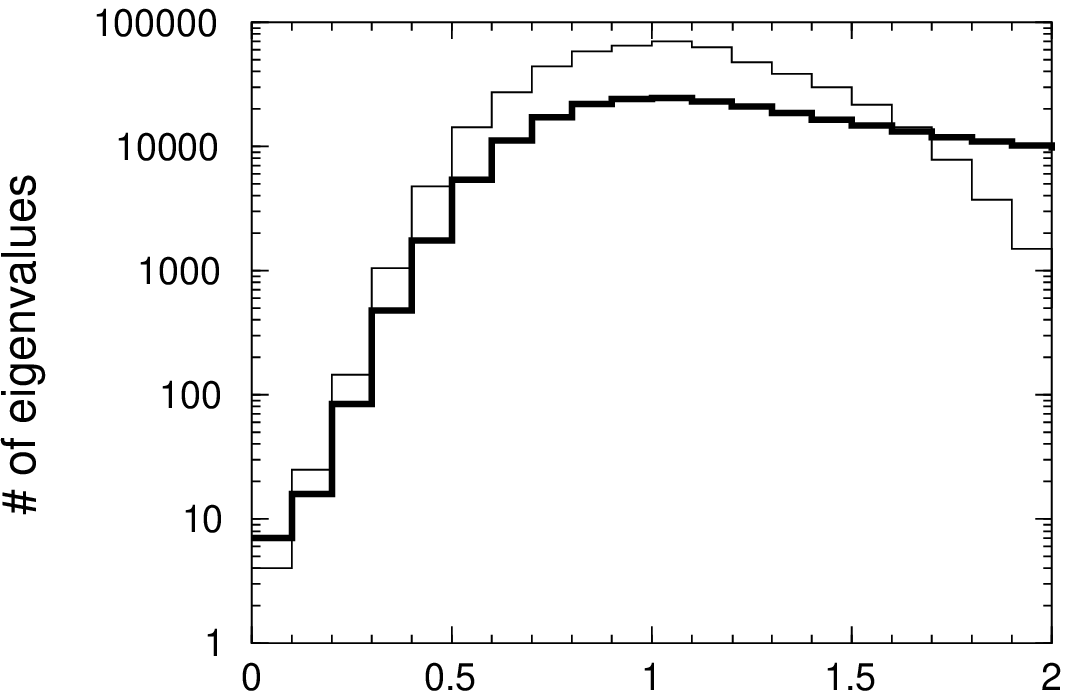} &
\hspace{-4mm}
\def\fpsangle{0} \epsfxsize=70mm \fpsbox{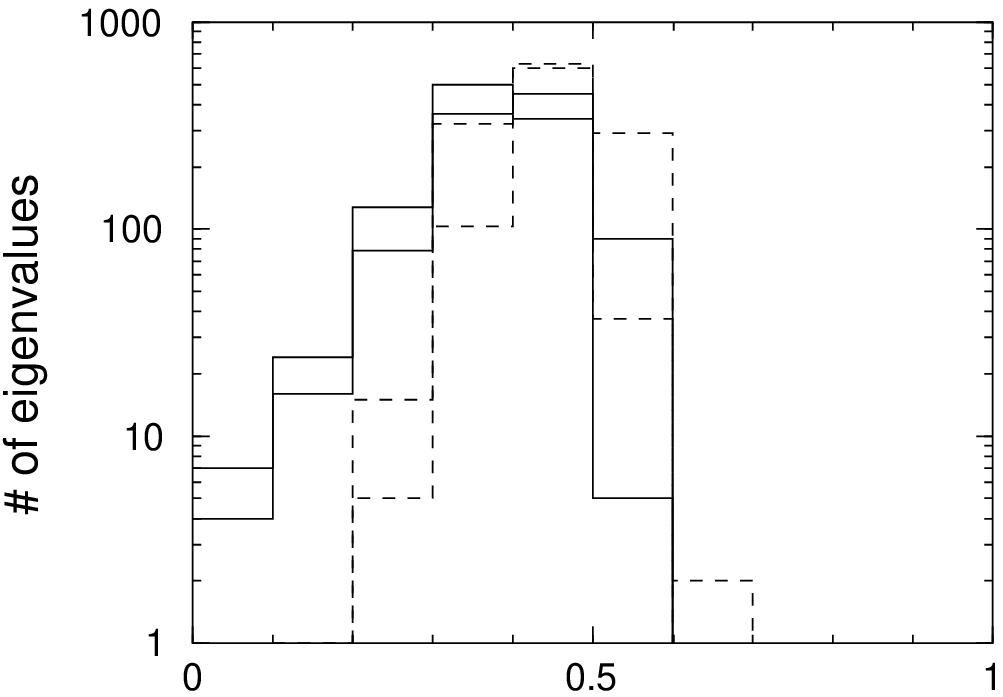}
\end{tabular}
\caption{\sl The eigenvalue distribution of $A^{\dagger}A$
at $\beta =6$ for the Neuberger fermion (bold) and the overlap
SO-HF (thin) from 1000 configurations. Left: all eigenvalues.
Right: the lowest two eigenvalues (solid and dashed).}
\label{figEV}
\end{figure}

A possible danger for the locality of an overlap operator
could still occur at strong coupling, if the eigenvalues of 
$A^{\dagger}A$ cluster very densely close to zero. %\cite{HJL}.
Since the use of an exact GW fermion (with respect to
$R_{x,y} = \delta_{x,y}/(2\mu )$) for $D_{0}$ fixes
$A^{\dagger}A = \mu^{2} = const.$ (for any configuration),
one could have hoped that an approximate GW fermion
suppresses the density of eigenvalues near zero.
However, from the eigenvalue histograms we could not
find any significant difference between the Neuberger fermion 
and the overlap SO-HF with this respect at $\beta =6$, and at
stronger coupling neither \cite{SCpri}.
In Fig.\ \ref{figEV} we show histograms of eigenvalues
obtained from 1000 configurations at $\beta =6$: the total
eigenvalue density (left) and the lowest two eigenvalues (right).
Also if we measure the separation of those lowest two eigenvalues
in each configuration, there is no significant difference between
the Neuberger fermion and the overlap SO-HF.
(Apparently the different degree of locality can be recognized
from the small eigenvalue distribution only if one takes into
account the (topological) quality of the considered eigenvalues.)

\begin{figure}[hbt]
%\hspace{-0.6cm}
\hspace*{25mm}
\def\fpsangle{0} \epsfxsize=80mm 
\fpsbox{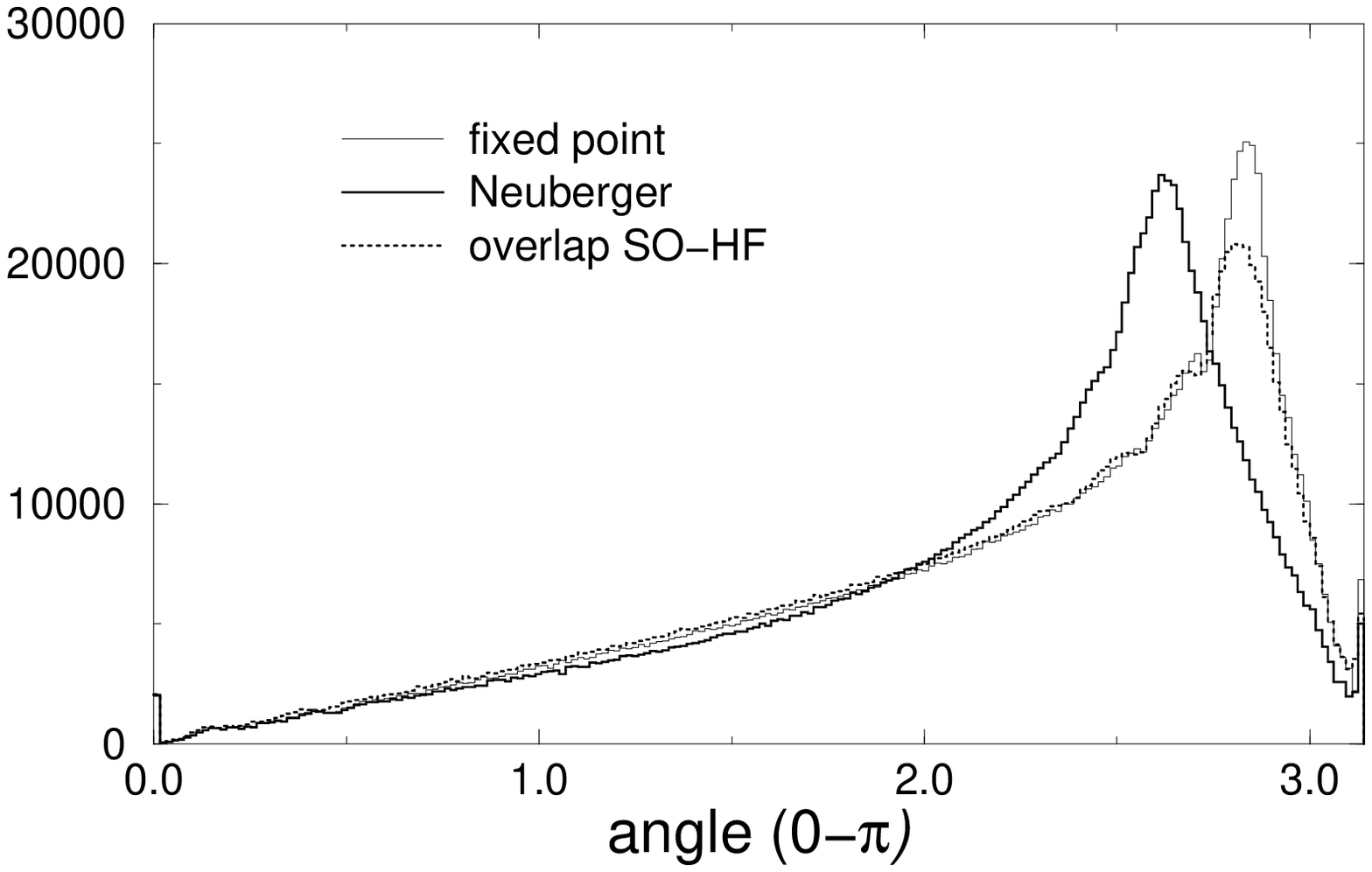}
\vspace{-3mm}
\caption{\sl The angular eigenvalue distribution on the unit
circle for various GW fermions, measured in 5000 quenched
configurations.}
\label{figang}
\end{figure}

As a last comparison, we show the angular density of the
overlap SO-HP and the Neuberger fermion (at $\beta =6$,
using $\mu=1$) compared to the FPA in Fig.\ \ref{figang}.
\footnote{For the Neuberger fermion, such an angular density
was studied before in Ref.\ \cite{SC}.}
Since there is a reflection symmetry on the real axis, we only show
the upper part, ranging from 0 ($\theta =0$) to 2 ($\theta = \pi $).
At small angles, i.e.\ small momenta, there is hardly any difference.
Again a clear distinction occurs at the opposite arc.
The overlap SO-HF has a peak at practically the same angle as the
fixed point fermion, in contrast to the Neuberger fermion.\\

To summarize this Section, we first repeat that the overlap-HFs are
exact GW fermions. If they scale very well in the straight application
and they approximate the GWR decently --- like the SO-HF ---
then the resulting overlap fermions do still scale well
{\em and} they have an exact remnant chiral symmetry.
So they combine two very important properties,
which were simultaneously present only in the FPA so far.
It's draw-back, however, is that its simulation is still
quite involved 
--- although the action is much simpler than the FPA.
In $d=4$ the square root operator is generally problematic,
and in addition we use a matrix $A$, which is
not as sparse as it is the case for the Neuberger fermion.
Therefore the application of such a formulation
in QCD might be rather expensive.
In the next Section we present a suggestion on how to
reduce the computational effort.

%1:1 mapping of the eigenvalues on the circle
%(for $D_{W}$ weird) \\
%meson dispersion, rotation invariance, LOCALITY,
%angular distribution

\section{Perturbative chiral correction}

We have seen in Section 2 that suitable HFs
can scale well and approximate the GWR reasonably well, up to
a certain coupling strength. In Section 3 we converted
these HFs into exact GW fermions by means of the overlap
formula, without a strong distortion of the good scaling
and rotation invariance. However, this formulation contains
an inconvenient square root operator. We now suggest a new method
to avoid that operator in order to reduce the computational
effort. For that purpose, we perform the 
``chiral projection'' perturbatively.

Assume that we start from an operator $D_{0}$, which is close
to a GW fermion for $R_{x,y} = \delta_{x,y}/(2\mu )$.
Then the operator
\begin{equation}
\varepsilon := A^{\dagger} A - \mu^{2}
\end{equation}
is small, $\Vert \varepsilon \Vert \ll 1$, and we use it as an
expansion term to approximate $(A^{\dagger}A)^{-1/2}$.
The perturbative chiral projection takes the form
\begin{eqnarray}
D_{pcp} &=& \mu - A \, Y \nonumber \\
O(\varepsilon ) &:& Y = \frac{1}{2} \, \Big[ \, 3 - 
\frac{1}{\mu^{2}} A^{\dagger} A \, \Big] \nonumber \\
O(\varepsilon^{2}) &:& Y = \frac{1}{8} \, \Big[ \, 15 - 
\frac{10}{\mu^{2}} A^{\dagger} A + \frac{3}{\mu^{4}} 
(A^{\dagger} A)^{2} \, \Big] \ , \quad {\rm etc.} \label{pertu}
\end{eqnarray}
One could write down the fully explicit action corresponding
to the Dirac operator $D_{pcp}$, but this form is complicated
already for $O(\varepsilon )$ (at least the absence of fat links
is profitable here).
However, for practical applications such an explicit form is
not needed, hence we do not write it down here.
Once the matrix-vector products $Ax$ and $A^{\dagger}x$
are implemented, the implementation of the $n^{\rm th}$
order chiral correction is trivial, and requires essentially
$1+2n$ such multiplications. Taking the first few orders is still
much cheaper than the exact chiral projection discussed in Section 3.
We emphasize that the linear growth in $n$ is very modest,
so it should be feasible to simulate this expansion also
beyond the lowest orders.

\begin{figure}[hbt]
\begin{tabular}{cc}
\hspace{-0.5cm}
\def\fpsangle{270} \epsfxsize=55mm \fpsbox{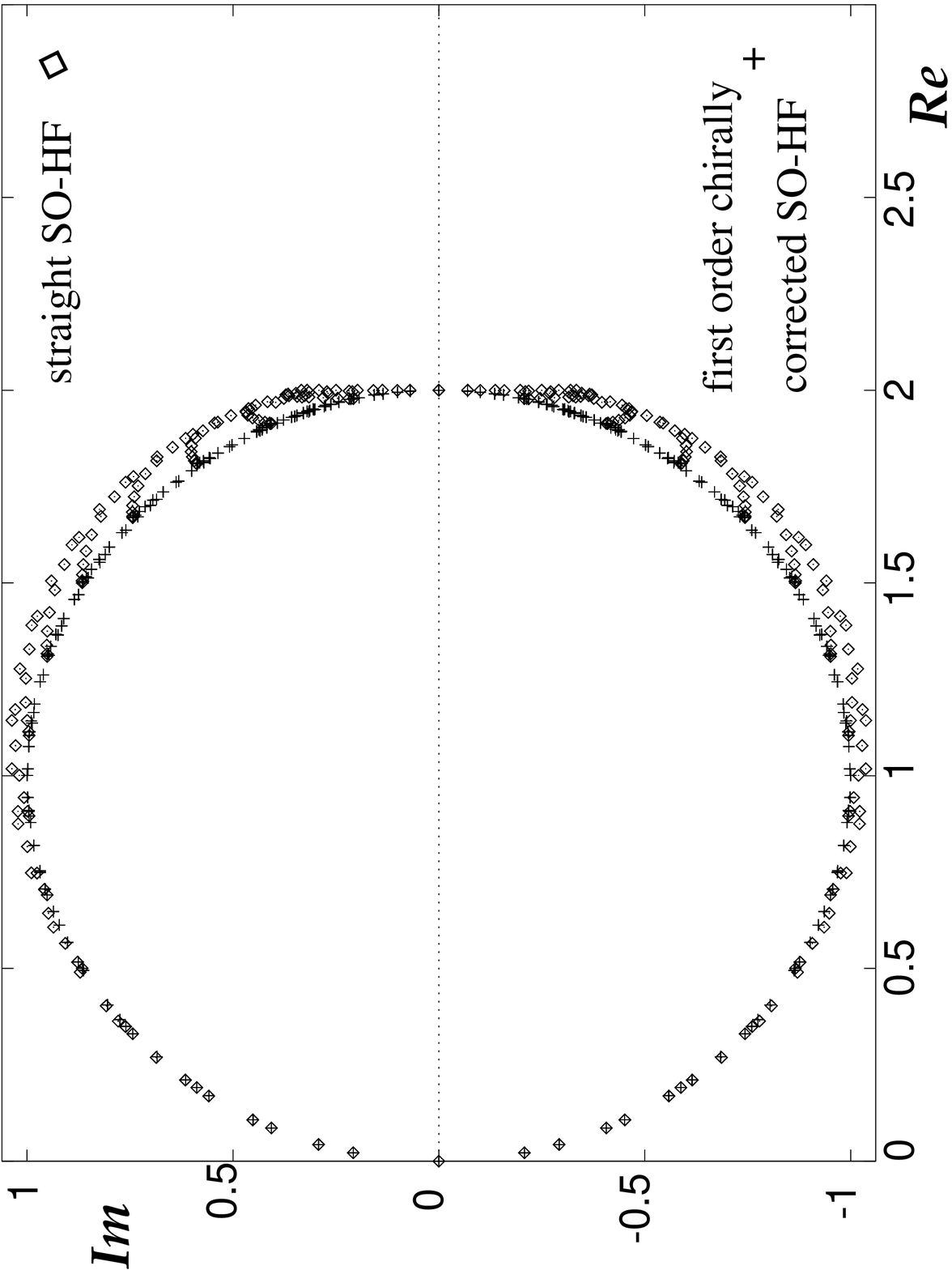} &
\hspace{-5mm}
\def\fpsangle{270} \epsfxsize=55mm \fpsbox{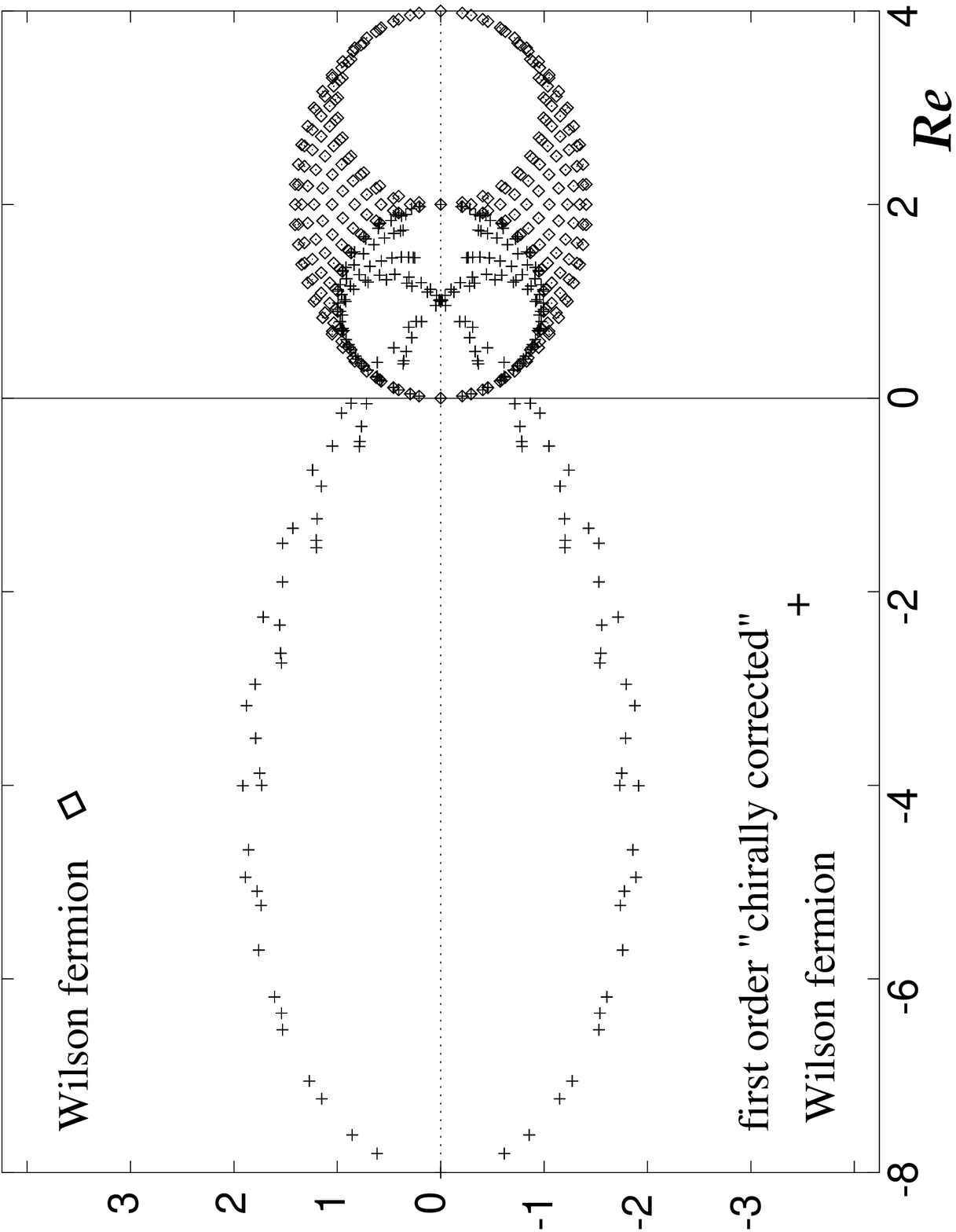}
\end{tabular}
\caption{\sl The initial and the first order chirally corrected 
free spectrum (using $\mu =1$)
for the SO-HF (left) and for the Wilson fermion (right)
on a $30 \times 30$ lattice.}
\label{figfreespecpert}
\end{figure}

Let us now discuss the efficiency of the perturbative
chiral projection. We consider the SO-HF, which scales
very well, but which violates the GWR most (among our HFs).
As a first example, we consider the spectrum of the free
SO-HF originally and after the first order chiral projection
for $\mu =1$. Fig.\ \ref{figfreespecpert} (left) 
shows that this first order does most of the projection
already; the resulting spectrum can hardly be distinguished from
the unit circle. 
%(although this is not even the closest circle 
%from the original spectrum).

Of course this method only works if we really start from an
approximate GW fermion.
\footnote{Although simple, the expansion of the square root
has not been used before, since all previous overlap simulations
used the Neuberger fermion.}
As an illustration, we show the spectrum
of the free Wilson fermion before and after first order ``chiral
projection'' in Fig.\ \ref{figfreespecpert} (right). 
Many eigenvalues do cluster at the unit
circle (those corresponding to small eigenvalues of $\varepsilon$),
but others diverge in this inadequate expansion
(those corresponding to eigenvalues of $\varepsilon$ with absolute
values $>1$). However, for even orders the ``tail of the comet''
flips far to the right-hand side, and perhaps it does not disturb
for practical purposes.

\begin{figure}[hbt]
\begin{tabular}{cc}
\hspace{-0.5cm}
\def\fpsangle{270} \epsfxsize=55mm \fpsbox{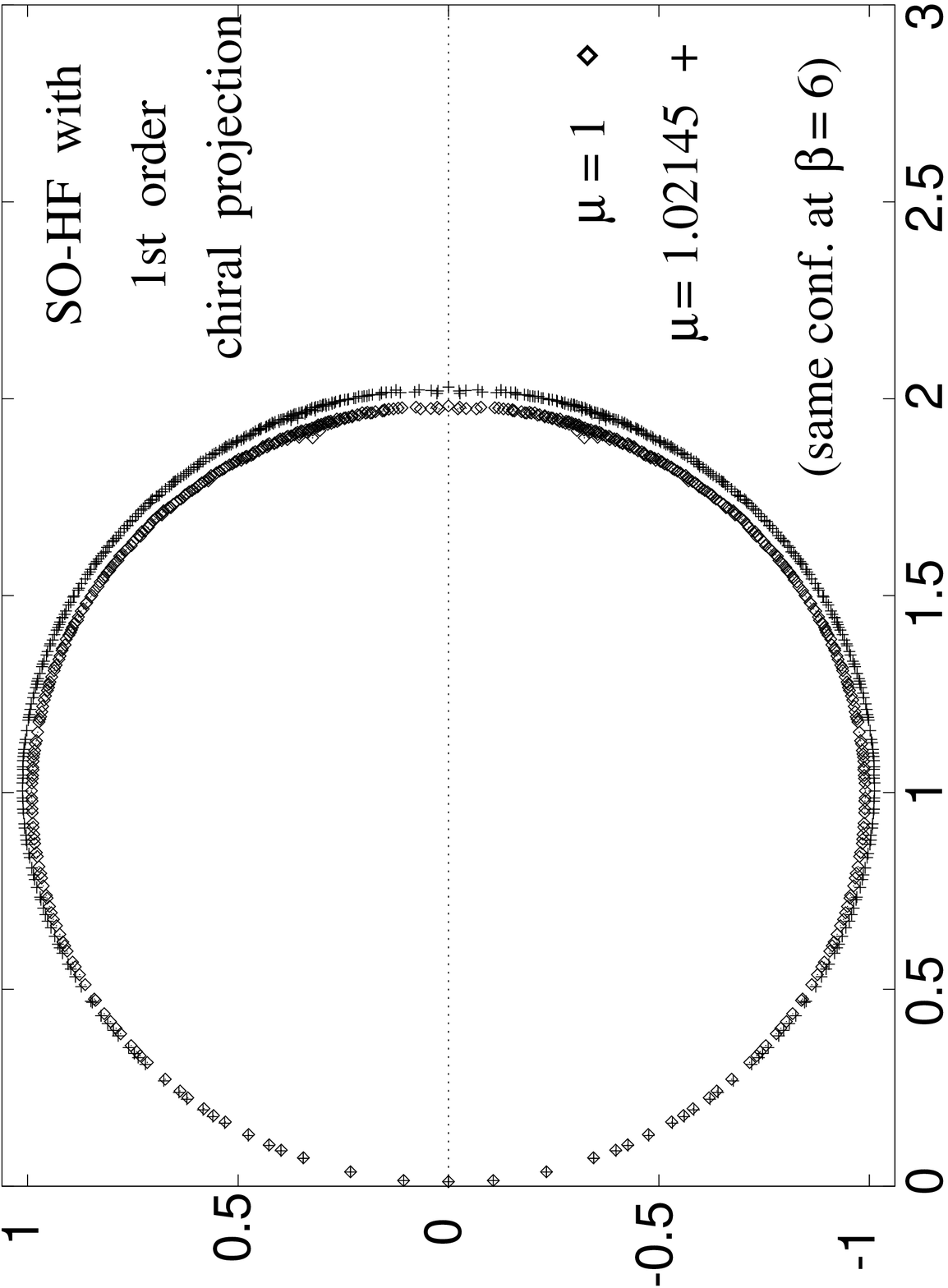} &
%\hspace{-2mm}
\def\fpsangle{0} \epsfxsize=70mm \fpsbox{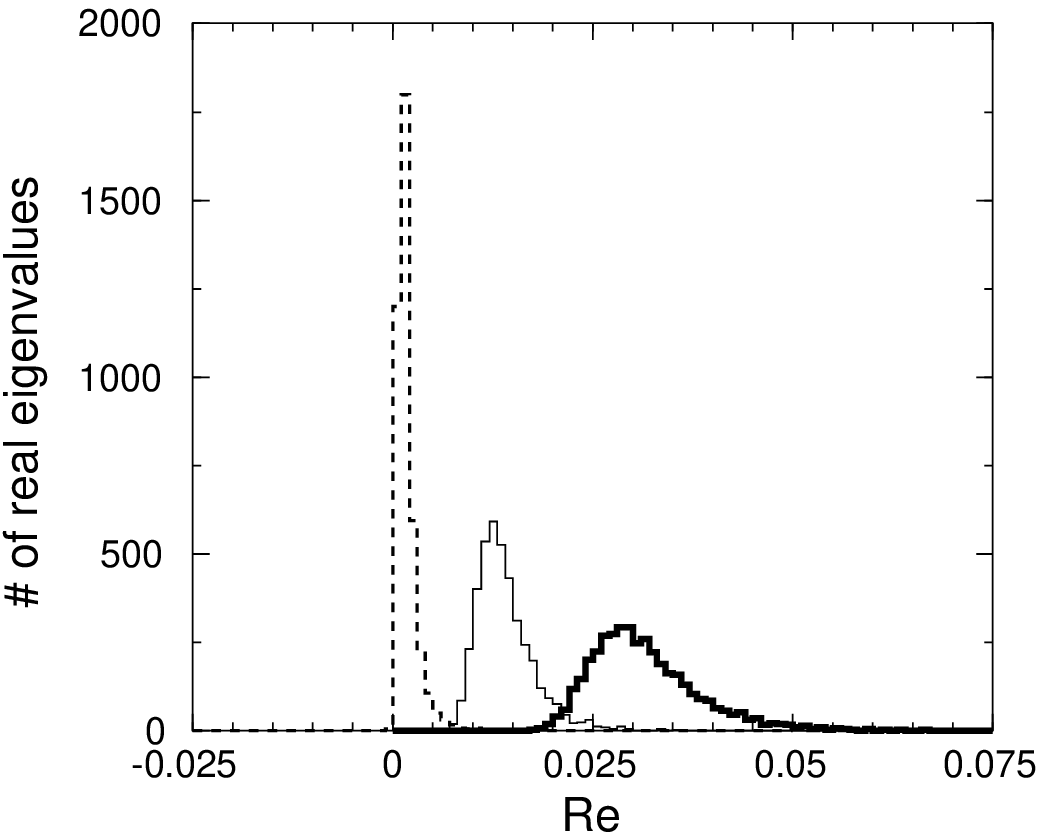}
\end{tabular}
\caption{\sl Left: The spectrum of the first order chirally
corrected SO-HF for a configuration at $\beta=6$ using
$\mu =1$ (smaller circle) resp. $\mu = 1.02145$ (larger circle).
Right: A histogram for the small, real eigenvalues 
of the SO-HF in 5000 quenched configurations at $\beta =6$:
initial (bold), first order chiral correction using $\mu =1.02145$
(solid) and the corresponding second order correction (dashed).
The corrections move the peak --- and hence the mass renormalization 
$\Delta m_{q}$ --- drastically towards zero
(see also Table \ref{tabDmd2}).}
\label{fighisto}
\end{figure}

We proceed to the Schwinger model, and we observe that --- for our
HFs at $\beta =6$ ---  the first order chiral correction does almost
the full projection already. 
As an example we consider the SO-HF. Its initial and fully
projected spectrum was shown for a typical configuration at
$\beta =6$ in Fig.\ \ref{figSO6-SO6_ov}.
We now show the first order mapping of the same configuration
to the unit circle, and to the closest circle (of radius $1.02145$)
in Fig.\ \ref{fighisto} (left). 
The radius of the larger circle corresponds to the optimal
value of $r_{0}$, which was identified in Section 2.1.
We observe in particular that the additive
``quark'' mass renormalization is pressed down form 
$\Delta m_{q} \simeq 0.032$ to about $0.014$ (and to $0.002$
for the second order), which corresponds to a 
reduction of $m_{\pi}$ from $0.13$ to $0.07$ (resp. $0.02$).
That effect is specifically illustrated in
Fig.\ \ref{fighisto} (right),
which shows the distribution of the small real eigenvalues
of 5000 configurations initially and after the perturbative
chiral correction to the first and to the second order.

%\begin{figure}[hbt]
%\hspace{-0.6cm}
%\hspace*{25mm}
%\def\fpsangle{0} \epsfxsize=80mm 
%\fpsbox{rehisto.eps}
%\vspace{-3mm}
%\caption{A histogram for the small, real eigenvalues 
%of the SO-HF in 5000 quenched configurations at $\beta =6$:
%initial (bold), first order chiral correction using $\mu =1.02145$
%(solid) and the corresponding second order correction (dashed).
%The peak --- and hence the mass renormalization $\Delta m_{q}$
%moves visibly to zero.}
%\label{fighisto}
%\end{figure}

\begin{figure}[hbt]
\begin{tabular}{cc}
\hspace{-0.5cm}
\def\fpsangle{270} \epsfxsize=55mm \fpsbox{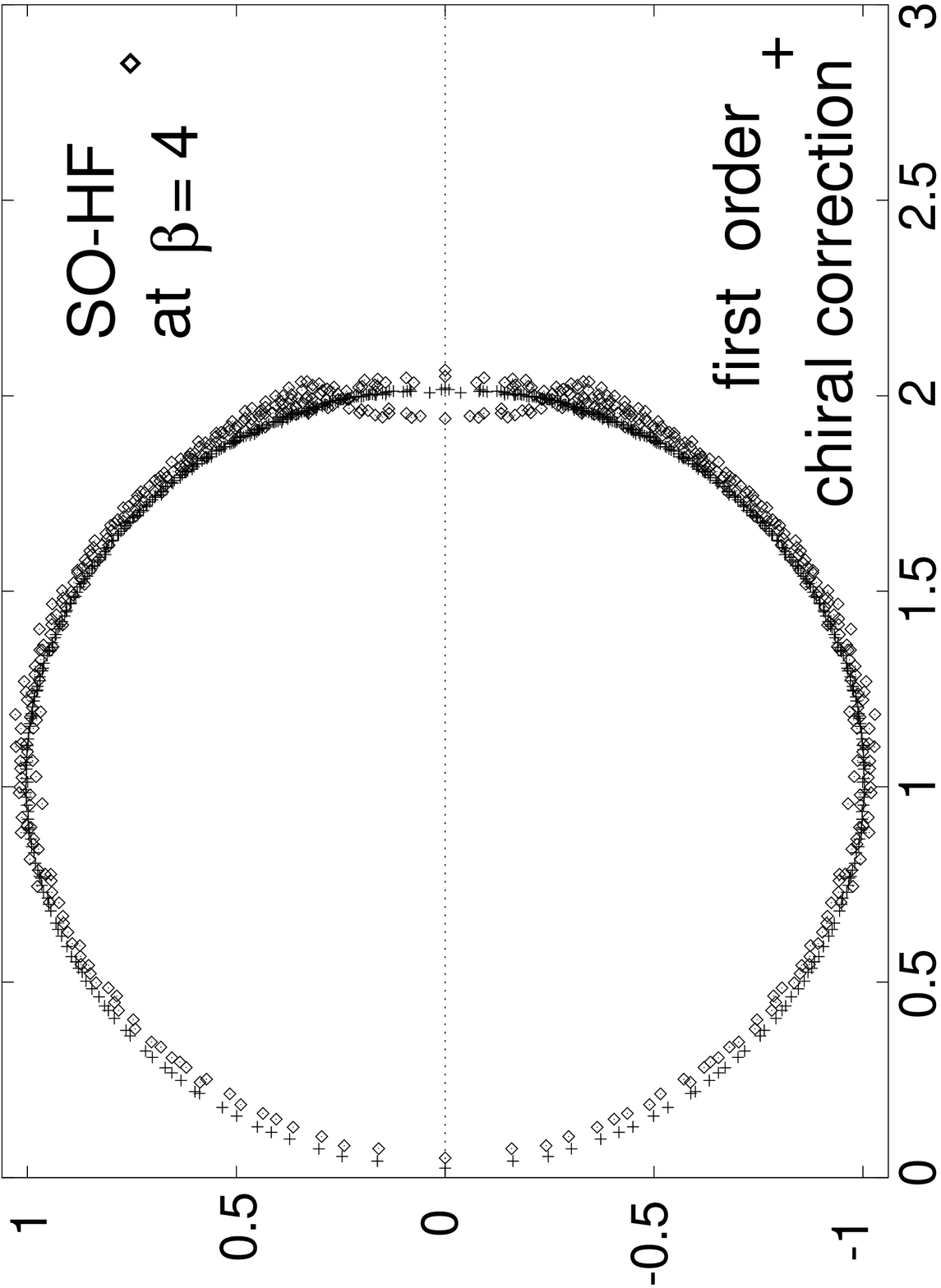} &
\hspace{-5mm}
\def\fpsangle{270} \epsfxsize=55mm \fpsbox{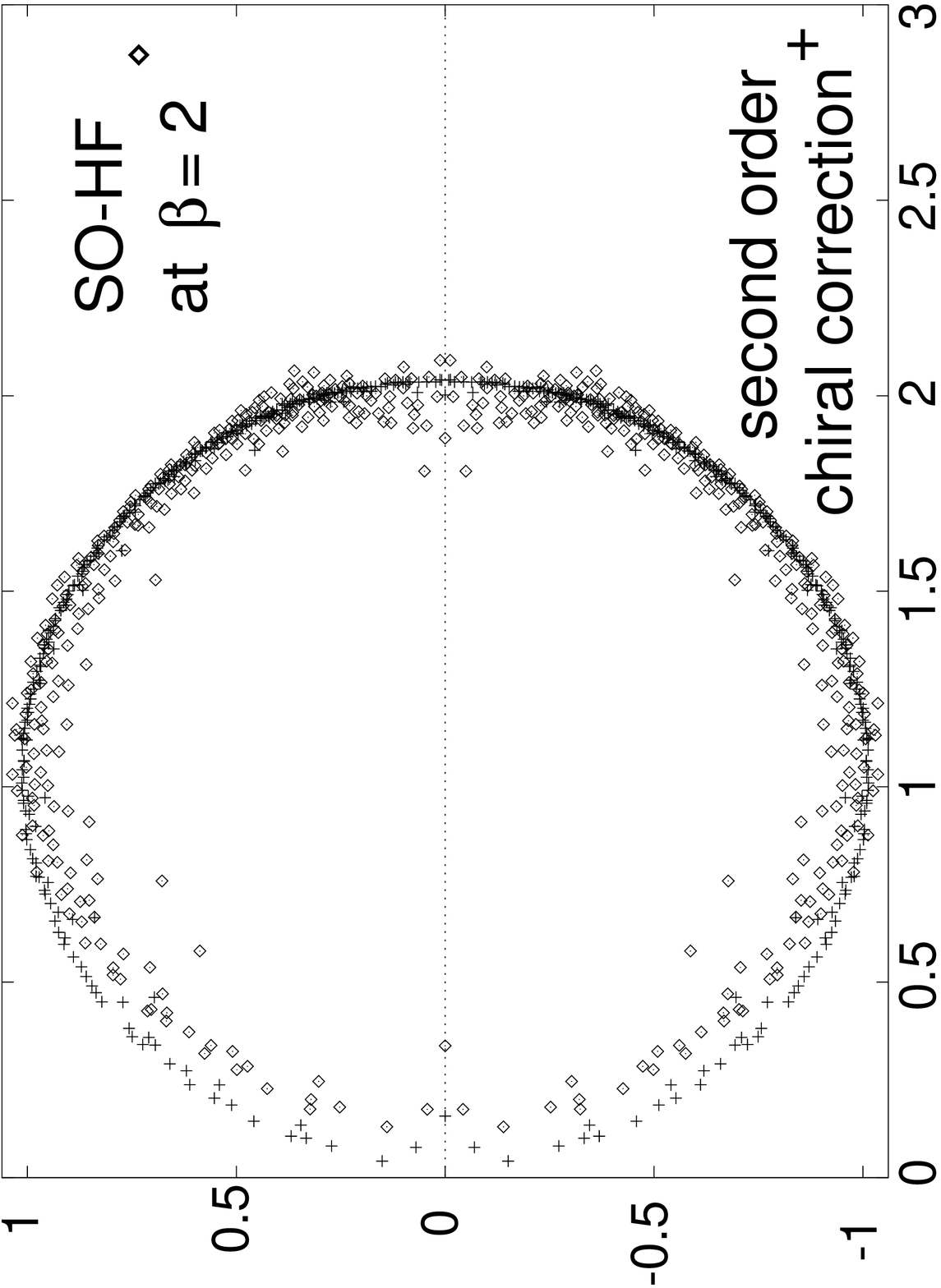}
\end{tabular}
\caption{\sl The spectra the SO-HF initially and after perturbative
chiral correction (using $\mu = 1.02145$): 
$\beta =4 $ and first order correction (left);
$\beta =2$ and second order correction (right).}
\label{figSO42}
\end{figure}

From Fig.\ \ref{figSO42} (left)
we see that at $\beta =4$ the first order correction is still efficient,
and finally Fig.\ \ref{figSO42} (right) shows that 
the second order can handle even strong coupling 
($\beta =2$) quite successfully. To quantify these observations
we present again the characteristic parameters $\Delta m_{q}$
and $\langle \delta^{2}_{r} \rangle$, which were introduced 
in Section 2.2, see Table \ref{tabDmd2}.

\begin{table}
{\small
\begin{center}
\begin{tabular}{|c|c|c|c|}
\hline
\hline
order of the  & & & \\ chiral correction & $\beta$ & 
$\Delta m_{q}$ & $\langle \delta_{r}^{2}\rangle$ 
(in units of $10^{-3}$) \\
\hline
\hline
   0  &  $\infty$ (free) &    0    &     0.444   \\
\hline
   1  &  $\infty$ &    0    &     0.000939   \\
\hline
   2  &  $\infty$ &    0    &     0.00000305   \\
\hline
\hline
   0  &     6     &    0.032(8)   &    0.48(5)   \\
\hline
   1  &     6     &    0.014(4)   &    0.21(4)   \\
\hline
   2  &     6     &    0.002(2)   &    0.0015(19)   \\
\hline
\hline
   0  &     4     &    0.066(67)   &    0.89(24)   \\
\hline
   1  &     4     &    0.036(58)   &    0.52(19)   \\
\hline
   2  &     4     &    0.0014(53)  &    0.02(12)   \\
\hline
\hline
   0  &     2     &    0.31(22)    &    5.9(20)    \\
\hline
   1  &     2     &    0.23(21)    &    4.4(17)   \\
\hline
   2  &     2     &    0.17(21)   &     1.3(11)   \\
\hline
\hline
 $\infty$ &  any  &      0      &      0    \\
\hline
\hline
\end{tabular}
\end{center}
}
\caption{\sl Two characteristic quantities to measure the
approximation of the standard GWR: the ``quark'' mass renormalization
$\Delta m_{q}$, and the mean value of the squared radial deviations
of the eigenvalues from the unit circle (in the complex plane),
$\langle \delta^{2}_{r} \rangle$ (as in Table \ref{tabstap}).
We show results for the SO-HF with chiral corrections using 
$\mu = 1.02145$.}
\label{tabDmd2}
\vspace{-3mm}
\end{table}

We have seen in Section 3 that the scaling and the
approximate rotation invariance are not badly affected by the
full chiral projection. In further tests we
made the very plausible observation
that the effect of a partial chiral projection is in between.
It corresponds roughly to the same interpolation as we just
observed in the spectrum.
Since the chiral behavior can be controlled by this economic
method, we would recommend its use most of all for the
SO-HF, which provides very good scaling from the beginning.

%\vspace*{5mm}

%concept: avoid square root, no need for explicit form of the 
%induced action.\\
%efficiency of first order in the free case and $\beta =6$;\\
%$\beta =4$ and 2 might require second order.
%Recommend chirally corrected SO-HF

%% [Is $(A^{\dagger}A)^{2}$ problematic ??] \ \ CPU comparison

%% [Actually we could also correct the SO-HF with respect
%% to $R_{x,y}=r_{0} \delta_{x,y}$, where $r_{0}$ is the optimal value
%% 0.4895. Then the correction should be even more powerful.]

\section{Conclusions}

Over the last year, the Ginsparg-Wilson relation became
fashionable in the lattice community; about
40 papers have been written about it. However, most of
the literature focuses on the chiral properties 
(and recently on algorithmic questions for
the Neuberger fermion \cite{algo,SCRI}) only.
Here we discussed chirality together with other crucial properties
of fermionic lattice actions, in particular the quality
of the scaling behavior, the approximate rotation invariance,
the locality and --- last but not least --- computational simplicity.

We presented three different approaches:
\begin{itemize}

\item In Section 2 we discussed the straight construction of approximate
GW fermions inside a short range. A few couplings
allow for a decent approximation. Such fermions are numerically
tractable in QCD, and in the Schwinger model we observe good scaling
behavior. However, at increasing coupling the GWR violation
becomes worse, which is manifest e.g.\ in a stronger mass
renormalization.

\item In Section 3 we corrected the GWR in the approximations
of Section 2 by means of the overlap formula.
We still observe good scaling and approximate
rotation invariance for the {\em improved overlap fermions}, 
in contrast to the usual Neuberger fermion.
However, this formulation is already somewhat demanding,
and it is expensive to apply it in $d=4$.

\item In Section 4 we show a way to simplify
the evaluation of the overlap action, by avoiding the tedious
square root operator. If we start from an approximate GW fermion,
then the square root can be replaced by a simple perturbative expansion,
and the first one or two orders are computationally relatively
cheap. They do, however, provide most of the chiral projection,
i.e.\ the GWR violation becomes small up to a considerable 
coupling strength. At the same time, scaling and rotation invariance
remain strongly improved, in particular for the ``scaling optimized
hypercube fermion'' SO-HF, which appears therefore as most 
satisfactory.

\end{itemize}

Based on these results, we think that an extension of this 
study to $d=4$ is highly motivated. 
Regarding the free fermion, all our results are
already relevant for $d=4$ too, because they correspond
to the special case $p_{3}=p_{4}=0$.
An issue in 4d gauge theory is for instance the choice of
a suitable action for the pure gauge part.\\

% For the first time in the world history, we have combined all
% favorable properties in one lattice action. We therefore
% expect to win the Nobel prize.\\

%\vspace*{2mm}

{\bf Acknowledgment} \ \ {\it We are indebted to S.\ Chandrasekharan,
who contributed to this work in an early stage. 
We thank him also for sharing with us his interesting ideas.
Furthermore we are very thankful to C.\ B.\ Lang for many 
important explanations. We also thank T. DeGrand, 
Ph.\ de Forcrand, F.\ Farchioni, M.\ L\"{u}scher, K.\ Orginos
T.\ Pany and K.\ Splittorff for inspiring comments. \
I.\ H.\ gratefully acknowledges his support
by the Fonds zur F\"{o}rderung der
Wissenschaftlichen Forschung in \"{O}sterreich,
Project P11502-PHY, and W.\ B.\ thanks for the kind hospitality
during his visit to Graz University, where this work picked
up its crucial momentum.}

\appendix

\section{Improved overlap fermions with and without mass}

We start this appendix by adding some more details about
the (massless) overlap fermions constructed from a hypercube
fermion $D_{HF} = \rho_{\mu} \gamma_{\mu} + \lambda$.
If we replace the momentum component $p_{2}$ by $-iE$,
then condition (1) in (\ref{cond12}) yields the free dispersion
relation
\begin{eqnarray} \nonumber
\cosh E &=& \frac{2 \rho^{(1)} \rho^{(2)} f(p_{1}) 
\pm \sqrt{1 + [ \rho^{(1) \, 2} - 4 \rho^{(2) \, 2} ] 
f(p_{1})}} {1-4 \rho^{(2) \, 2} f(p_{1})} \ , \quad {\rm with}\\
f(p_{1}) &:=& \left( \frac{\sin p_{1}}{\rho^{(1)} + 2 \rho^{(2)} 
\cos p_{1}} \right) ^{2} .
\end{eqnarray}
In all the cases considered in Section 3.1 the lower sign does not
lead to a real energy $E$, hence there are no upper branches.
\footnote{Also in all further free dispersions in this appendix,
the unphysical sign never yields a real energy.}
According to condition (2) this curve stops as soon as
\begin{equation}
\lambda_{0} + 2 \lambda_{1} ( \cos p_{1} + \cosh E)
+ 4 \lambda_{2} \cos p_{1} \cosh E \geq \mu \ .
\end{equation}
For $D_{W}$ with mass $M$, this simplifies to
\ $\cosh E = \sqrt{1+\sin^{2}p_{1}}$ \ , ending at
\ $\cos p_{1} + \sqrt{1+\sin^{2}p_{1}} \geq 2+M-\mu$.
We observed in Section 3 that the choice $\mu \approx 1$ sets the 
end-point to a useful position (for $M=0$).

Taking into account the normalization, the shape of the curve depends on
one single parameter, say $\rho^{(2)}$. 
In the center of the Brillouin zone, the curve rises monotonously
with $\rho^{(2)}$, and a good interpolation between $D_{W}$
($\rho^{(2)}=0$) and $D_{TP-HF}$ ($\rho^{(2)}=0.0953$) makes it 
practically coincide with the continuum dispersion up to
$p_{1} \simeq \pi /2$. In this way, the SO-HF was constructed,
with a scalar term close to our other two HFs.
We saw that it scales excellently, also in the straight application
(without overlap).

\begin{figure}[hbt]
\begin{tabular}{cc}
\hspace{-0.5cm}
\def\fpsangle{270} \epsfxsize=55mm \fpsbox{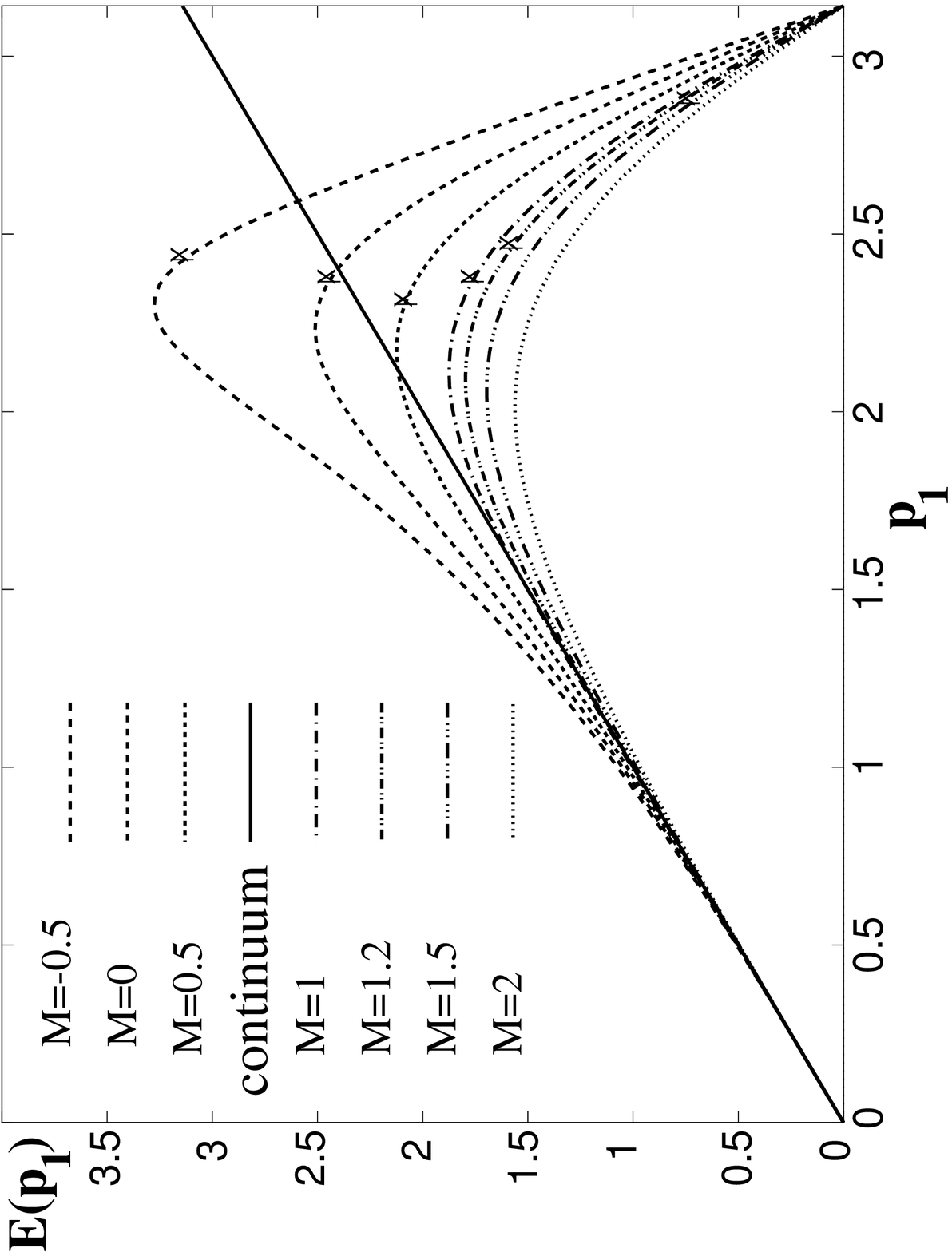} &
\hspace{-5mm}
\def\fpsangle{270} \epsfxsize=55mm \fpsbox{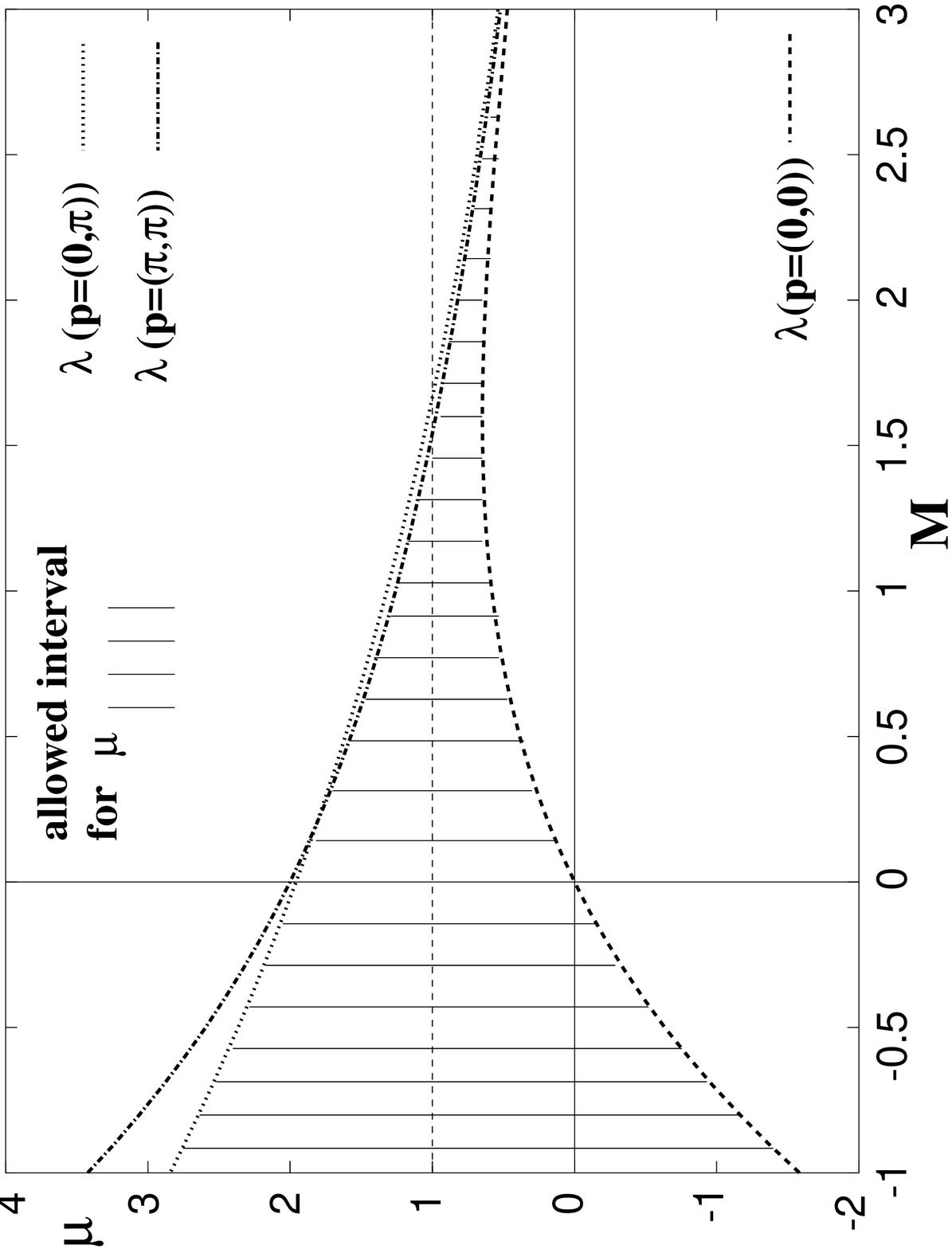}
\end{tabular}
\caption{\sl Left: The free dispersion relation for overlap fermions
constructed from massive TP-HFs (with mass parameter $M$).
Again we mark the end-points for $\mu =1$.
Right: The allowed region for $\mu$ to provide one flavor for
overlap TP-HFs, depending on $M$.}
\label{figGWfreedispM}
\end{figure}

Nothing prevents us from also {\em inserting massive HFs in the overlap
formula}; we still obtain massless GW fermions. 
As an example, we use the massive TP-HF \cite{MIT}, 
where the parameter $M$ is the mass of the continuum fermion, 
which was originally blocked to the lattice.
\footnote{For the TP-HF this implies a bare mass of
$M^{2} / (\exp (M) -1)$.}
If we vary $M$, the dispersion curve of the overlap fermion
can again be deformed monotonously, and a value close to $M=1$
appears optimal, see Fig.\ \ref{figGWfreedispM} (left).
The overlap fermion constructed from the TP-HF at $M\approx 1$
performs also very well in the thermodynamic tests described
in Sections 2 and 3, on the same level as the SO-HF. 
However, in this region of $M$ the 1-species range of 
the mass parameter $\mu$ --- introduced in Section 3 --- is 
very narrow, as we see from Fig.\ \ref{figGWfreedispM} (right).
\footnote{The asymmetry with respect to the sign of $M$ arises
from the kernel in the Gaussian blocking term. We obtain
optimal locality for $R^{-1}_{x,y}(M) = \pm \delta_{x,y}
M^{2}/(\exp (M)-M-1)$. Above we have chosen the positive sign.
Then the couplings of the TP-HF at $M=1$ are: $\rho^{(1)}=0.11163921$,
$\rho^{(2)}=0.02885462$, $\lambda_{0}=1.10520159$, 
$\lambda_{1}=-0.09226400$, $\lambda_{2}=-0.03854222$.
For the negative sign in $R^{-1}$: $\lambda_{i} \to - \lambda_{i}$.}
Since this is dangerous for the interacting
case, we consider the SO-HF as a better approach.\\

Finally we add some remarks on the case, where an overlap fermion
is made massive by adding a mass term at the end.
Such massive fermions are likely to have still very good chiral
properties, which is useful for simulations of heavy quarks.

\begin{figure}[hbt]
\begin{tabular}{cc}
\hspace{-0.5cm}
\def\fpsangle{270} \epsfxsize=55mm \fpsbox{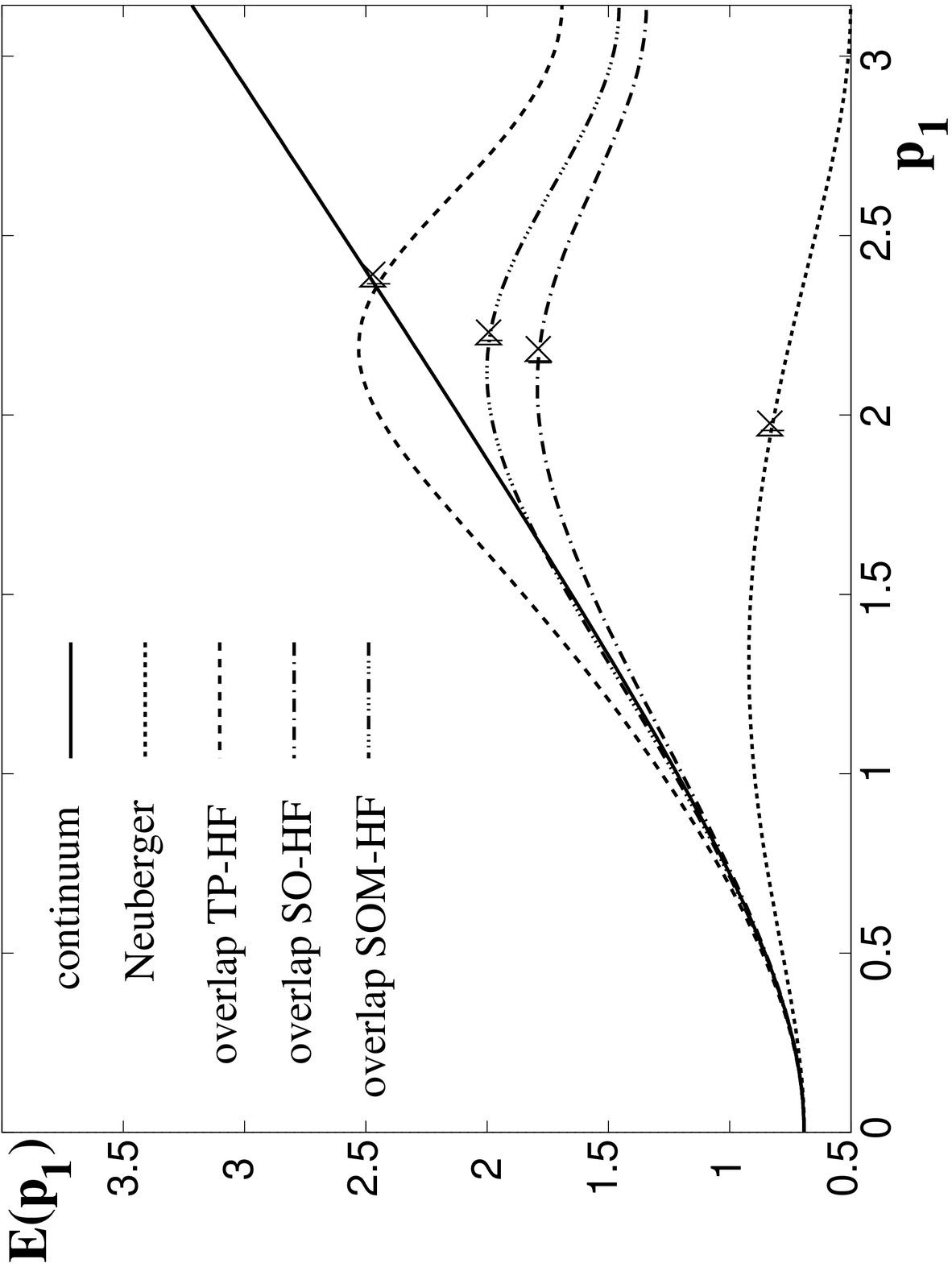} &
\hspace{-4mm}
\def\fpsangle{270} \epsfxsize=55mm \fpsbox{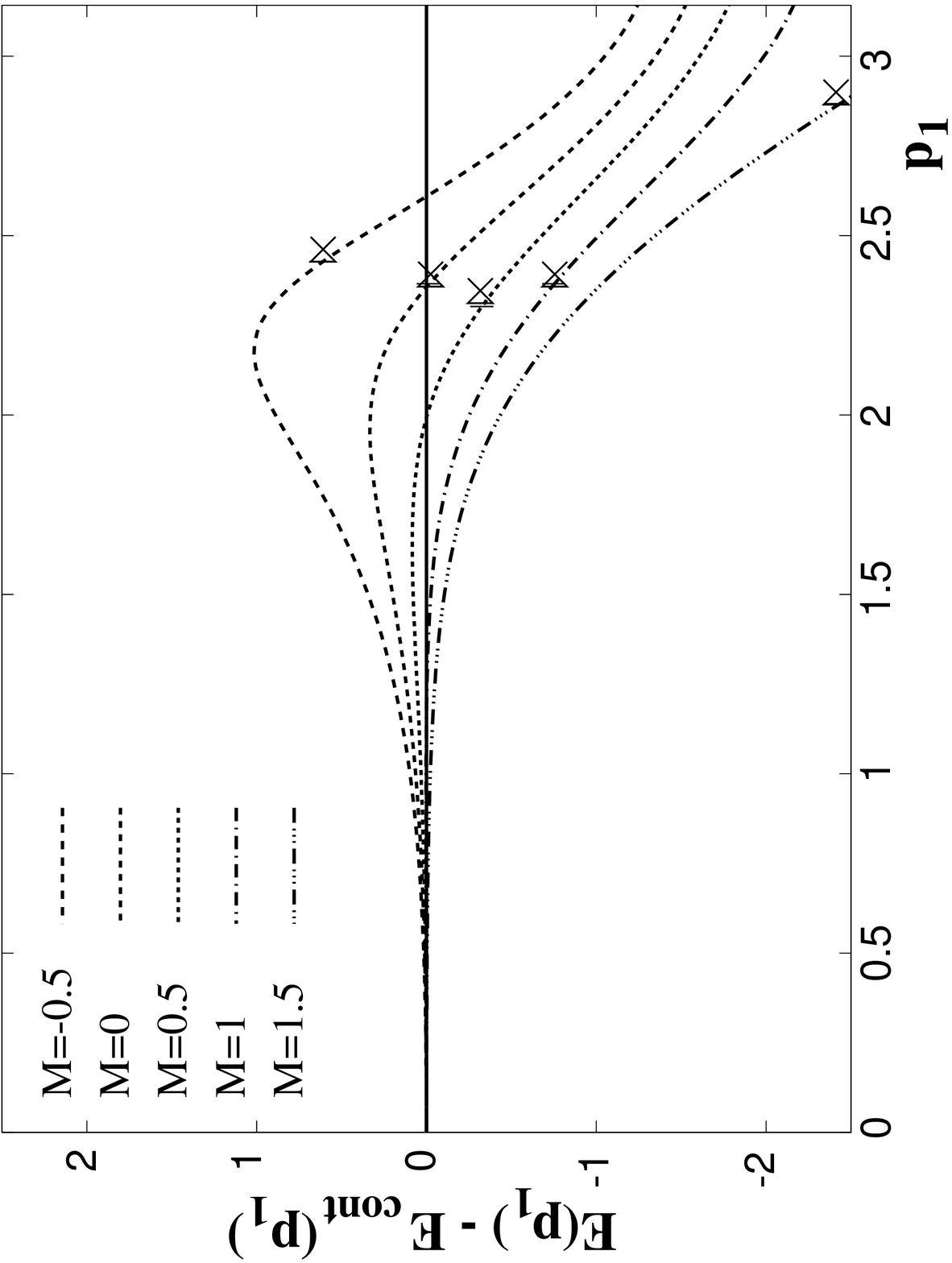}
\end{tabular}
\caption{\sl Left: The free dispersion relation for massive
overlap fermions constructed from $D_{W}$ or from massless HFs.
Right: The free dispersion relation for massive
overlap fermions constructed from massive HFs. In both plots we
mark again the end-points for $\mu =1$, as a useful example.}
\label{figmGW}
\end{figure}

We add the mass term in the straightforward manner,
\footnote{
T.-W.\ Chiu suggested an alternative way \cite{Chiu}. 
In the notation of Section 1 it amounts to
%\begin{displaymath}
%D'_{\tilde m} = \frac{D_{\chi} + \tilde m}{1 + RD_{\chi}} \ .
%\end{displaymath}
$ D'_{\tilde m} = (D_{\chi} + \tilde m)/(1 + RD_{\chi})$ . 
However, if $R_{x,y} \propto \delta_{x,y}$ then this is equivalent
to eq.\ (\ref{mass}), due to $D_{m} = (1+Rm)D'_{\tilde m}$,
if $\tilde m = m /(1+Rm)$, 
as S.\ Chandrasekharan first noticed \cite{SCpri}. }
\begin{equation} \label{mass}
D_{m} = 1 + m + \frac{A}{\sqrt{A^{\dagger} A}} \ .
\end{equation}
If we now search for poles in the free $D^{-1}(p)$, then
condition (1) is generalized compared to (\ref{cond12}),
\begin{equation}
(1) \qquad \rho^{2} = \bar m \ (\lambda - \mu )^{2} \ ,
\quad \bar m := 1 - \frac{4}{(1+m + \frac{1}{1+m} )^{2} } \ ,
\end{equation}
while condition (2) keeps the same form.
In Fig.\ \ref{figmGW} (left) 
we show the free dispersion relations for various
massive overlap fermions. 
(We now omit the CO-HF; its scaling is always slightly
worse than the TP-HF.)
They are all constructed by using a massless
$D_{0}$ and adding $m=1$. We see that the overlap SO-HF
is still good here, but not optimal any more. If we increase
$\rho^{(2)}$ a little to $0.088$, then we obtain the
SOM-HF, which scales excellently in this case.
\footnote{For completeness we give the full set of couplings of the
SOM-HF: $\rho^{(1)}=0.324$, $\rho^{(2)}=0.088$; $\lambda_{0}=1.5$,
$\lambda_{1}=-0.25$, $\lambda_{2}=-0.125$.}
There is no reason for the optimal value of $\rho^{(2)}$
not to depend on $m$.

\begin{figure}[hbt]
\begin{tabular}{cc}
\hspace{-0.5cm}
\def\fpsangle{270} \epsfxsize=55mm \fpsbox{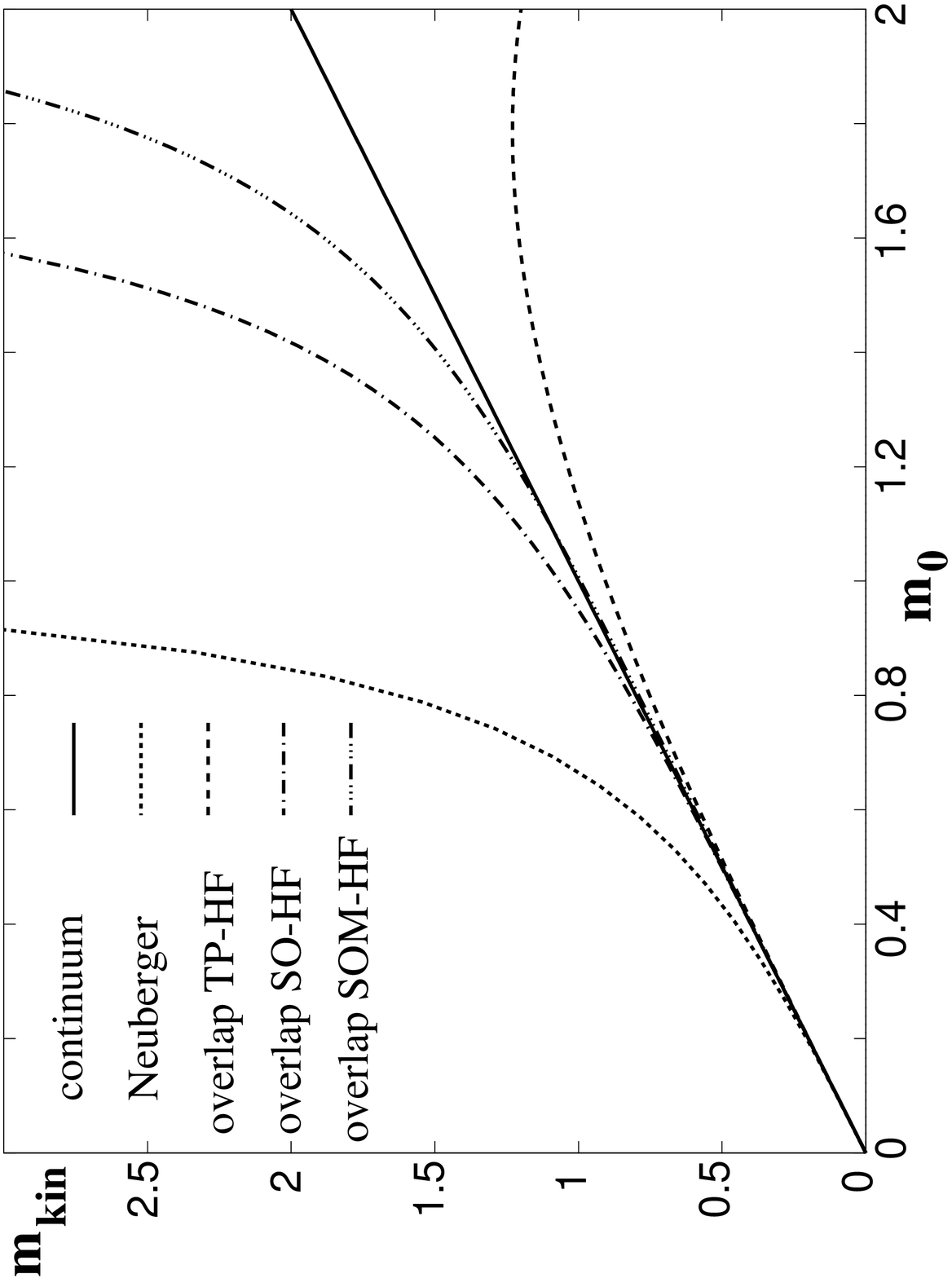} &
\hspace{-5mm}
\def\fpsangle{270} \epsfxsize=55mm \fpsbox{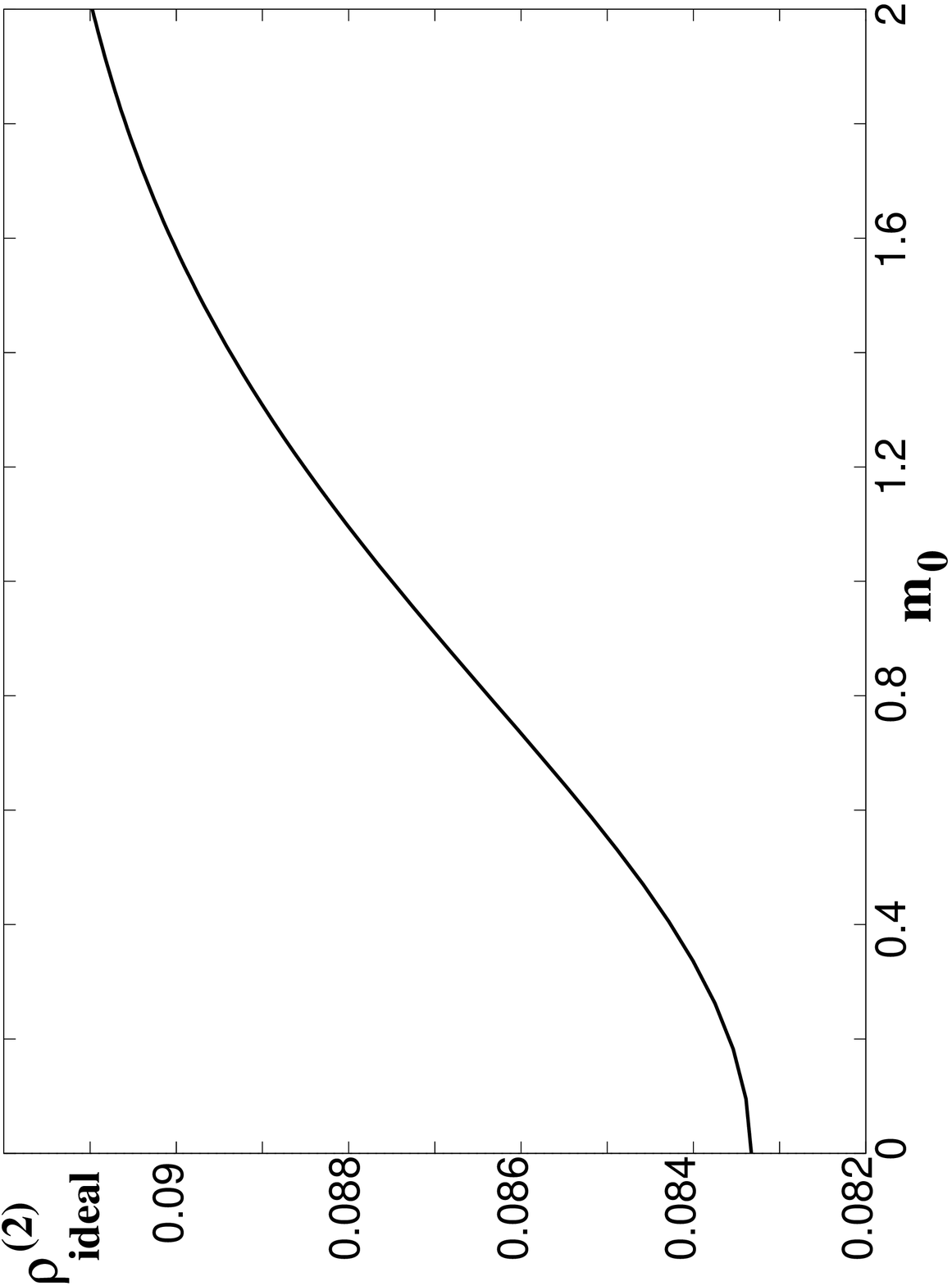}
\end{tabular}
\caption{\sl Left: Kinetic mass $m_{kin}$ vs.\ static mass $m_{0}$
for various types of overlap fermions.
Right: The coupling $\rho^{(2)}_{ideal}$ in the massive, scaling
optimized HF, which provides $m_{0}=m_{kin}$ after insertion
in the overlap formula.}
\label{figm0_mkin}
\end{figure}

The low momentum expansion of the massive overlap operator starts with
\begin{eqnarray}
D_{m}(p) &=& m_{0} + \frac{1}{2 m_{kin}} p^{2} + O(p^{4}) \ , \nonumber \\
m_{0} &=& {\rm arcosh} \frac{1}{\sqrt{1 - \bar m}} \ , \nonumber \\
m_{kin} &=& \frac{\sqrt{\bar m}}{4} \Big(
\rho^{(1)\, 2} - \rho^{(2)} + \frac{8 \rho^{(1)} \rho^{(2)} + 
\bar m \, \lambda_{1}}{2 \sqrt{1 - \bar m}} +
\frac{ \rho^{(2)} + 4 \rho^{(2)\, 2} + \bar m \, \lambda_{2}}{1 -\bar m}
\Big)^{-1} .
\end{eqnarray} 
This confirms for instance the value \ $m_{0}(m=1) \simeq 0.693$ \ 
for all the massless fermions considered in Fig.\ \ref{figmGW}.
In Fig.\ \ref{figm0_mkin} (left)
we plot the static vs.\ kinetic mass for 
various overlap fermions, and we see again that the overlap
SO-HF performs well at small $m_{0}$. If $m$ is of order 1, 
however, the two masses agree better for the overlap SOM-HF.
In a next step, we fix the $\lambda$ term of the
SO-HF ($\lambda_{1}=2\lambda_{2}=-1/4$), and we tune the remaining free
parameter $\rho^{(2)}$ to an ``ideal value'', so that $m_{kin}=m_{0}$. 
The resulting $\rho^{(2)}_{ideal}(m_{0})$ is shown in
Fig.\ \ref{figm0_mkin} (right).
It reveals a smooth mass dependence of the scaling optimal overlap HF.
\footnote{Our value for $\rho^{(2)}$ in the SOM-HF is a little larger
than $\rho^{(2)}_{ideal}(0.693)$, because this helps the dispersion
to follow the continuum curve closely up to larger momenta.}

Finally we remark that we can also insert a massive $D_{0}$,
and make the resulting overlap fermion massive again.
As an example, we consider some massive TP-HFs with varying
mass parameters $M$, which are all converted to a massive
overlap fermion by adding $m=1$. Here the static masses
$m_{0}$ come out differently depending on $M$.
For a direct comparison we show in Fig.\ \ref{figmGW}
(right) the difference between the energy $E(p_{1})$
and the continuum energy $E_{cont}(p_{1})$.
Again the vicinity of $M=1$ looks very good, but --- as we mentioned
before --- that fermion might be too close to zero- or multi-species 
(doubling) to be useful in gauge theory.

%free case only, fermion dispersion, $D_{HF}$ with mass m, 
%SO, SOM, $m_{0}=m_{kin}$

\end{document}